%% file: nonabelian.tex
\newif\iffigs\figstrue

\documentstyle[12pt]{article}
\iffigs
  \input epsf
\else
\fi

\batchmode
\newfont{\footscrfont}{rsfs10}
  \newfont{\footbbbfont}{msbm10}
  \newfont{\manfont}{manfnt}
\errorstopmode

\newif\ifscrf\scrftrue
\ifx\footscrfont\nullfont
  \scrffalse
\fi

\newif\ifamsf\amsftrue
\ifx\footbbbfont\nullfont
  \amsffalse
\fi

\def\ppnumber{\vbox{\baselineskip14pt\hbox{CU-TP 924}
\hbox{hep-th/9811201}}}
\def\ppdate{November 1998}
\def\pplogo{\vbox{\kern-\headheight\kern -15pt
\halign{##&##\hfil\cr&{
\ppnumber}\cr\rule{0pt}{2.5ex}&\ppdate\cr}
}}

\makeatletter
\date{}
\def\dedicatory#1{\def\@date{\normalsize\it#1}}
\def\subjclass#1{\def\@thefnmark{}\@footnotetext{1991
    {\it Mathematics Subject Classification.} #1}}
\def\keywords#1{\def\@thefnmark{}\@footnotetext{
    {\it Key words and phrases.} #1}}

\def\ps@firstpage{\ps@empty \def\@oddhead{\hss\pplogo}%
  \let\@evenhead\@oddhead 
}
\def\maketitle{\par
 \begingroup
 \def\thefootnote{\fnsymbol{footnote}}
 \def\@makefnmark{\hbox
 to 0pt{$^{\@thefnmark}$\hss}}
 \if@twocolumn
 \twocolumn[\@maketitle]
 \else \newpage
 \global\@topnum\z@ \@maketitle \fi\thispagestyle{firstpage}\@thanks
 \endgroup
 \setcounter{footnote}{0}
 \let\maketitle\relax
 \let\@maketitle\relax
 \gdef\@thanks{}\gdef\@author{}\gdef\@title{}\let\thanks\relax}

\def\abstract{\if@twocolumn
\section*{Abstract}
\else \small
\begin{center}
{\bf ABSTRACT}
\end{center}
\quotation
\fi}

\def\thebibliography#1{\section*{References\@mkboth
 {REFERENCES}{REFERENCES}}\small\list
 {[\arabic{enumi}]}{\settowidth\labelwidth{[#1]}\leftmargin\labelwidth
 \advance\leftmargin\labelsep
 \usecounter{enumi}}
 \def\newblock{\hskip .11em plus .33em minus .07em}
 \sloppy\clubpenalty4000\widowpenalty4000
 \sfcode`\.=1000\relax}

\newif\iffn\fnfalse

\@ifundefined{reset@font}{\let\reset@font\empty}{} 
\long\def\@footnotetext#1{\insert\footins{\reset@font\footnotesize
    \interlinepenalty\interfootnotelinepenalty
    \splittopskip\footnotesep
    \splitmaxdepth \dp\strutbox \floatingpenalty \@MM
    \hsize\columnwidth \@parboxrestore
   \edef\@currentlabel{\csname p@footnote\endcsname\@thefnmark}\@makefntext
    {\rule{\z@}{\footnotesep}\ignorespaces
      \fntrue#1\fnfalse\strut}}}

\makeatother

\ifamsf
  \newfont{\bigbbbfont}{msbm10 scaled\magstep2}
  \newfont{\bbbfont}{msbm10 scaled\magstep1}  
  \newfont{\smallbbbfont}{msbm8}
  \newfont{\tinybbbfont}{msbm6}
  \newfont{\smallfootbbbfont}{msbm7}
  \newfont{\tinyfootbbbfont}{msbm5}
  \newfont{\biggthfont}{eufm10 scaled\magstep2}
  \newfont{\gthfont}{eufm10 scaled\magstep1}  
  \newfont{\smallgthfont}{eufm8}
  \newfont{\tinygthfont}{eufm6}
  \newfont{\footgthfont}{eufm10}
  \newfont{\smallfootgthfont}{eufm7}
  \newfont{\tinyfootgthfont}{eufm5}
\fi

\ifscrf
  \newfont{\scrfont}{rsfs10 scaled\magstep1}  
  \newfont{\smallscrfont}{rsfs7}
  \newfont{\tinyscrfont}{rsfs7}
  \newfont{\smallfootscrfont}{rsfs7}
  \newfont{\tinyfootscrfont}{rsfs7}
\fi

\ifamsf
  \newcommand{\Bbb}[1]{\iffn
      \mathchoice{\mbox{\footbbbfont #1}}{\mbox{\footbbbfont #1}}
      {\mbox{\smallfootbbbfont #1}}{\mbox{\tinyfootbbbfont #1}}\else
      \mathchoice{\mbox{\bbbfont #1}}{\mbox{\bbbfont #1}}
      {\mbox{\smallbbbfont #1}}{\mbox{\tinybbbfont #1}}\fi}
  
\else
  \def\bigbbbfont{\bf}
  \def\Bbb{\bf}
  
\fi

\ifscrf
  \newcommand{\Scr}[1]{\iffn
    \mathchoice{\mbox{\footscrfont #1}}{\mbox{\footscrfont #1}}
    {\mbox{\smallfootscrfont #1}}{\mbox{\tinyfootscrfont #1}}\else
    \mathchoice{\mbox{\scrfont #1}}{\mbox{\scrfont #1}}
    {\mbox{\smallscrfont #1}}{\mbox{\tinyscrfont #1}}\fi}
\else
  \def\Scr{\cal}
\fi

\def\tablerule{\noalign{\hrule}}
\def\operatorname#1{\mathop{\rm #1}\nolimits}
\def\C{{\Bbb C}}

\def\Q{{\Bbb Q}}
\def\R{{\Bbb R}}
\def\Z{{\Bbb Z}}

\def\bearray{\begin{eqnarray}}
\def\eearray{\end{eqnarray}}
\def\bearraynn{\begin{eqnarray*}}
\def\eearraynn{\end{eqnarray*}}
\def\bfig{\begin{figure}}
\def\efig{\end{figure}}

\def\diag{\operatorname{diag}}
\def\opeq#1{\advance\lineskip#1 \advance\baselineskip#1
	\advance\lineskiplimit#1}

\def\cM{{\Scr M}}

\def\cD{{\Scr D}}

\def\cMc{{\hfuzz=100cm\hbox to 0pt{$\;\overline{\phantom{X}}$}\cM}}
\def\barcD{{\hfuzz=100cm\hbox to 0pt{$\;\overline{\phantom{X}}$}\cD}}

\ifamsf

\else

\fi

\newtheorem{Proposition}{Proposition}[section]

\newtheorem{Theorem}{Theorem}[section]
\newtheorem{Lemma}{Lemma}[section]
\newtheorem{Corrolary}{Corrolary}[section]

\newcommand{\be}{\begin{equation}}
\newcommand{\ee}{\end{equation}}
\newcommand{\bea}{\begin{eqnarray}}
\newcommand{\eea}{\end{eqnarray}}

\newcommand{\bp}{\begin{Proposition}}
\newcommand{\ep}{\end{Proposition}}
\newcommand{\bt}{\begin{Theorem}}
\newcommand{\et}{\end{Theorem}}
\newcommand{\bl}{\begin{Lemma}}
\newcommand{\el}{\end{Lemma}}
\newcommand{\bc}{\begin{Corrolary}}
\newcommand{\ec}{\end{Corrolary}}
\newcommand{\nn}{\nonumber}

\begin{document}

\title{D-branes on Nonabelian Threefold Quotient Singularities}

\author{Brian~R.~Greene$^{1,a}$,\\
C.~I.~Lazaroiu$^{2,b}$ and Mark Raugas$^{3,c}$ }

\maketitle
\vbox{
\centerline{$^1$Departments of Physics and Mathematics}
\centerline{$^2$Department of Physics}
\centerline{$^3$Department of Mathematics}
\centerline{Columbia University}
\centerline{N.Y., N.Y. 10027}
\medskip
\medskip
\bigskip
}

\abstract{We investigate the classical moduli space of $D$-branes on 
a {\em nonabelian} Calabi-Yau threefold singularity and find that 
it admits topology-changing transitions. We construct a general formalism
of worldvolume field theories in the language of quivers and give a 
procedure for computing the enlarged K\"{a}hler cone of the moduli space. 
The topology changing transitions achieved by varying the 
Fayet-Iliopoulos parameters correspond to changes of 
linearization of a geometric invariant theory quotient and can 
be studied by methods of algebraic geometry.  Quite surprisingly, 
the structure of the enlarged K\"{a}hler cone can be computed by toric 
methods. 
By using this approach, we give a detailed discussion of two low-rank 
examples.}

\vskip .6in
$^a$ greene@phys.columbia.edu

$^b$ lazaroiu@phys.columbia.edu

$^c$ raugasm@math.columbia.edu

\pagebreak

\section*{Introduction}

The purpose of the present paper is to explore some new aspects of 
{\em D-geometry}, the geometry of ultrashort distances in string theory 
as seen by D-branes. 
As first pointed out in \cite{Shenker} and discussed in more detail in 
\cite{DKPS} from a D-brane point of view \cite{Polchinski}, 
the presence of non-perturbative objects 
implies the ability of string theory to probe distances below the string 
scale. Given the fascinating geometric phenomena known to occur in the moduli 
space of perturbative string theory (see, for example, \cite{topchange}), 
which have changed our 
understanding of the way in which this theory modifies the 
general-relativistic concepts of space and time, it is natural to ask whether 
such processes survive, and in what form, when nonperturbative string physics 
is included. A first step towards a satisfactory answer to this 
question is to consider the behaviour of D-branes located at an orbifold 
singularity of a Calabi-Yau space, and to ask how D-brane physics reflects
the underlying geometry of space-time. This approach is similar in spirit 
to studies of perturbative string theory on orbifold singularities 
undertaken during the past decade \cite{orbifolds} 
(see also the review \cite{aspin}), 
which gave the first indications of how  
string theory modifies classical geometric concepts. 

Various steps in the direction mentioned above have been taken in a 
series of papers
\cite{branes2, branes3}, where it was shown that well-known phenomena of 
perturbative string theory on orbifolds (such as resolution of singularities) 
survive when D-branes are included, although the underlying physical 
mechanisms are somewhat different. In the case of quotient singularities in 
two complex dimensions, i.e. orbifold singularities locally of 
the form $\C^2/\Gamma$, with $\Gamma$ a finite subgroup of 
$SU(2)$, an almost complete study has been performed \cite{branes2} 
by using the powerful mathematical results of \cite{Kronheimer,KN,Nakajima}. 
These investigations lead to a very beautiful and natural  
realization of minimal resolutions of canonical surface singularities via physical processes in string theory. 

In the case of $D$-brane 
quotient singularities of Calabi-Yau threefolds (which have the 
local form $\C^3/\Gamma$ with $\Gamma$ a finite subgroup of $SU(3)$), the 
focus until now has been exclusively on the case when the group $\Gamma$ is 
abelian. In this case, the problem can be reduced \cite{branes3,Infirri} 
to one in toric geometry, for which powerful computational tools are available
\footnote{Toric geometry is a machinery for studying varieties 
$X$ which admit a densely-embedded complex torus, by systematically 
reducing all geometric questions about $X$ to questions in {\em integral} 
convex geometry. The computational power of this approach is belied by the 
existence of a rich variety of algorithms in computational convex geometry, 
a field which has attracted considerable interest during the last few 
decades due to its 
widespread applications to various areas of mathematics. While referring the 
interested reader to the physics-oriented introductions in 
\cite{topchange,toric_phys}
or to the mathematical account of \cite{toric}, we hurry to assure him or her
that the present paper can be read without having any previous 
knowledge of this topic.}. 
On the other hand, progress in the {\em nonabelian} case has been 
stifled by the 
lack of a comparably versatile mathematical tool. One of the main aims of this 
paper is to show that the problem is tractable (albeit difficult), and that, 
surprisingly enough, one can extract important information by using 
abelian (i.e. toric) techniques. This will be achieved by carefully 
identifying the information we consider most relevant, and by separating it 
from the context, which allows us to avoid the hard computational task of 
giving a complete characterization of the geometry. As
an example, by following 
this approach, we will be able to establish the existence of topology-changing 
transitions in the nonabelian 3-dimensional context.
In fact, we will show 
that some crucial features of the abelian case carry over to nonabelian 
situations, and that the phenomenon of topology change 
can be understood in these generalized settings.  
A more general purpose of the present work 
is to develop the formalism appropriate for describing the nonabelian case. 
As we will argue in this paper, the 
best formalism seems to be that of {\em quivers}, a subject which has received 
considerable attention recently in the mathematical literature
\cite{quivers_intro,quivers_moduli} concerned 
with singularity theory and the representation theory of algebras.
We explain the relevant concepts of this subject in a form adapted to our 
problem and in accesible language. As we will discover, 
quiver technology can be used to considerably simplify the analysis 
of D-brane effective field theories placed at quotient singularities.

Another motivation for this paper is the recent effort aimed at 
generalizing the AdS/CFT conjecture of \cite{AdS} to the case of D-branes 
on conical singularities of Calabi-Yau manifolds \cite{AdScon}. Such 
theories can be obtained \cite{MP} by turning on Fayet-Iliopoulos parameters 
in the effective field theory of a large number of branes placed at a 
quotient singularity, and then flowing to a conformal fixed point of the 
resulting field theory. The number of examples one can consider as candidates
for applying this method has been so far limited to abelian quotient 
singularities, due the lack of appropriate methods for treating the effective 
field theory in the nonabelian case. The present paper is a first step 
toward removing that constraint, thus preparing the way for more 
extensive studies of such conformal fixed points. In a sense, the situation
is similar to that of supersymmetric field theories in 4 dimensions, where 
a thorough understanding of the classical moduli space is a necessary 
pre-requisite for a quantum-mechanical study along the lines of 
\cite{Seiberg}.

The plan of this paper is as follows. In Section 1 we discuss the effective 
field theory describing the low energy dynamics of D-branes placed at a 
general Calabi-Yau threefold quotient singularity and we describe its 
classical moduli space of vacua. 
In Section 2, we use quiver techniques in order to simplify the problem of 
computing the moduli space. We show that the moduli space of vacua 
can be identified with the moduli space of representations of an 
associated quiver. This enables us to apply recent mathematical results to 
the problem of understanding topology-changing transitions, a subject we 
discuss in Section 3. 
In Section 4, we explain how one can extract 
information about such transitions by using toric methods. Finally, 
in Section 5 we apply these methods to two examples (obtained by taking 
$\Gamma$ to be a low rank finite subgroup of $SU(3)$), giving a detailed 
account of the arguments involved. Certain technical details are 
discussed in the appendices.

\section{The projection conditions and the moduli space}

In this section, as indicated above, we discuss the field
theory degrees of freedom necessary to describe D-branes
at a nonabelian quotient singularity   in three complex
dimensions. To do so, let's
consider a finite subgroup $\Gamma$ of order $N$ of 
$SU(3)$, acting on $\C^3$ by its 
defining representation (i.e. via its embedding as a subgroup of $SU(3)$). 
The model of one $Dp$-brane on a $\C^3/\Gamma$ orbifold, which  
meets the $D$-brane at a point, can be formulated by the methods of 
\cite{Polchinski}.
This approach starts with a $U(N)$ gauge 
theory with adjoint scalars ${\cal X}_a$~($a=1..6$)~ 
$({\cal X}_a^+={\cal X}_a)$, 
changing to complex coordinates:
\bea
X_m:={\cal X}_m+i{\cal X}_{m+3}~~\nn\\
X_{\overline m}:=X_m^+={\cal X}_m-i{\cal X}_{m+3}~~,
\eea
with $m=1..3$, and imposing the projection conditions: 
\bea
D^{(R)}(\gamma)X_mD^{(R)}(\gamma)^{-1}=D^{(Q)}_{nm}(\gamma)X_n~~\nn\\
D^{(R)}(\gamma)UD^{(R)}(\gamma)^{-1}=U~~,
\eea
on the complex fields $X_m$ and the gauge group elements $U$. Here 
$D^{(R)}$ is the regular
\footnote{More general representations can be chosen, with 
appropriate physical interpretations \cite{fractional}, but in this paper 
we consider the regular representation only.} 
$N\times N$ representation of $\Gamma$, 
%
%
while $D^{(Q)}(\gamma)=(D^{(Q)}_{nm}(\gamma))_{n,m=1..3}$ is the defining 
representation. The well-known decomposition 
$R=\oplus_{\mu=0..r}{n_\mu R_\mu}$ 
(where $R_\mu$ ($\mu=0..r$) are the irreducible 
representations  of $\Gamma$, with $R_0$ the trivial representation
and $n_\mu:=dim R_\mu$) shows that we can choose the basis of Chan-Paton 
factors such that the matrices $D^{(R)}(\gamma)$ have the form:
\be
D^{(R)}(\gamma)=\left[\begin{array}{ccccc}
D^{(0)}(\gamma)\otimes 1_{n_0} &0 &0  &...&0 \\
0& D^{(1)}(\gamma)\otimes 1_{n_1} &0 &...&0\\
0& 0 & D^{(2)}(\gamma)\otimes 1_{n_2}&...&0\\ 
...& ...&...& ...&...\\
0&0&0&...& D^{(r)}(\gamma)\otimes 1_{n_r}\\
\end{array}\right]~~,
\ee
where $D^{(\mu)}(\gamma)\otimes 1_{n_\mu}$ represents
\footnote{Throughout this paper, the symbol $1_m$ denotes the $m$ by $m$ 
identity matrix, or the identity map of an $m$-dimensional vector space.}
the $n_\mu^2$ by 
$n_\mu^2$ block-diagonal matrix which contains $n_\mu$ copies of 
$D^{(\mu)}(\gamma)$:
\be
D^{(\mu)}(\gamma)\otimes 1_{n_\mu}=\left[\begin{array}{cccc}
D^{(\mu)}(\gamma)&0&...&0\\
0&D^{(\mu)}(\gamma)&...&0\\
...&...&...&...\\
0&...&0&D^{(\mu)}(\gamma)\end{array}\right]~~.
\ee

%
%
The gauge group $G_0$ of the projected theory is the subgroup of $U(N)$ formed 
by those elements $U$ which satisfy the projection conditions. Applying 
Schur's lemma shows that such an element has the form:
\be
U=\left[\begin{array}{cccc}
1_{n_0}\otimes U_0&0&...&0\\
0&1_{n_1}\otimes U_1&...&0\\
...&...&...&...\\
0&...&0&1_{n_r}\otimes U_r\end{array}\right]~~,
\ee
with $U_\mu$ arbitrary $n_\mu$ by $n_\mu$ unitary matrices. Therefore, $G_0$ 
is isomorphic to $\Pi_{\mu=0..r}{U(n_\mu)}$. 

The scalar potential of the theory is given by:
\be
V=-\sum_{a,b=1..6}{Tr([{\cal X}_a,{\cal X}_b]^2)}~~,
\ee
which in complex variables becomes:
\be
V=\frac{1}{4}Tr\sum_{m,n=1..3}{([X_m,X_n][X_m,X_n]^+ +
[X_m,X_{\overline n}][X_m,X_{\overline n}]^+)}~~.
\ee
Using the identity:
\bea
Tr\sum_{m,n=1..3}{[X_m,X_{\overline n}][X_m,X_{\overline n}]^+} =
Tr\sum_{m,n=1..3}{[X_m,X_n][X_m,X_n]^+}+\nn\\
+Tr(\sum_{m=1..3}{[X_m,X_{\overline m}]})
(\sum_{m=1..3}{[X_m,X_{\overline m}]})^+ ~~,
\eea
this can be rewritten as:
\be
V=\frac{1}{2}[V_f+V_d]~~, 
\ee
where:
\bea
V_f:=Tr\sum_{m,n=1..3}{[X_m,X_n][X_m,X_n]^+}~~\nn\\
V_d:=\frac{1}{2}Tr(\sum_{m=1..3}{[X_m,X_{\overline m}]})
(\sum_{m=1..3}{[X_m,X_{\overline m}]})^+~~.
\eea
Introducing the moment map for the gauge group action:
\be
\label{moment1}
\rho=\sum_{m=1..3}{[X_m,X_{\overline m}]}~~,
\ee
we can write:
\be
V_d=\frac{1}{2}Tr(\rho^2)~~.
\ee
In the presence of Fayet-Iliopoulos terms, this relation is modified to:
\be
\label{zeta}
V_d=\frac{1}{2}Tr[(\rho-\zeta)^2]~~,
\ee
where 
\be
\zeta=\left[\begin{array}{cccc} 
\zeta_0 1_{n_0^2}&0&...&0\\
0&\zeta_1 1_{n_1^2}&...&0\\
...&...&...&...\\
0&...&0&\zeta_r 1_{n_r^2}\end{array}\right]~~,
\ee
is a matrix in the centre of the surviving gauge group\\ 
(note that $\zeta_\mu 1_{n_\mu^2}=\left[\begin{array}{cccc} 
\zeta_\mu 1_{n_\mu}&0&...&0\\
0&\zeta_\mu 1_{n_\mu}&...&0\\
...&...&...&...\\
0&...&0&\zeta_\mu 1_{n_\mu}\end{array}\right]$).
In this situation, the supersymmetric vacuum constraints are:
\bea
[X_m,X_n]=0~~\nn\\
\rho(X_1,X_2,X_3)=\zeta~~.
\eea
Equation (\ref{moment1}) enforces the constraint $Tr(\rho)=0$, i.e. 
$\sum_{\mu=0..r}{n_\mu^2\zeta_\mu}=0$. 
\footnote{This constraint is a feature of working with the regular 
representation. If one replaces $R$ by a more general representation, 
then the  expression 
of the moment map is different, and the diagonal $U(1)$ subgroup acts 
nontrivially on the fields present in that case.  In particular, the 
tracelessness condition ceases to hold.}. Dividing out the
diagonal $U(1)$ subgroup of $G_0$ 
(which acts trivially on $X_m,X_{\overline m}$) gives an effective action of 
$G=G_0/U(1)_{diag}$. 
Define the {\em variety of commuting matrices} ${\cal Z}$ to be the set of 
matrices satisfying the projection conditions, and the equations 
$[X_m,X_n]=0$ for all $m,n=1..3$. Then the desired moduli space is the 
K\"{a}hler quotient:
\be
{\cal M}_\zeta:=\{X \in {\cal Z}|\rho(X)=\zeta\}/G~~.
\ee

Notice that the variety of commuting matrices coincides with the set of 
extremal points of the function:
\be
\label{superpot}
W=\epsilon_{mnl}Tr(X_mX_nX_l)~~.
\ee
That is, the conditions $[X_m,X_n]=0$ for all $m,n=1..3$ are equivalent to the 
constraints $\partial_{X_m^{ij}}W=0, ~\forall i,j=1..N, ~\forall m=1..3$. 
If one is dealing with $D3$-branes, then the projected theory has $N=1$ 
supersymmetry in 4 dimensions and $W$ is its superpotential. In this case, 
the condition $V_f=0$, i.e. $[X_m,X_n]=0$ for all $m,n=1..3$ is simply 
the F-flatness constraint, while the condition $V_d=0$ is the D-flatness 
constraint in the Wess-Zumino gauge.  By analogy with that situation, 
we will in general call $V_d=0$ the {\em D-flatness} constraint and 
$V_f=0$ the ${\em F-flatness}$ constraint.

\section{Quiver formalism}

The above formulation is inconvenient because of 
the presence of linear constraints (imposed by the projection conditions) 
among the components of the matrices $X_m$.
One can avoid this complication by performing a linear change of 
variables which solves the projection conditions.  This 
amounts to parametrizing the space of projected matrices $X_m$ 
by a set of unconstrained variables $\phi$ which are free to fill some linear 
space (further constraints on $\phi$ will be later imposed by the F-flatness 
conditions).
In essence, the so-called quiver formalism is a procedure which
accomplishes this change of variables in a particularly
systematic and transparent manner. To see this,
view the triple $X=(X_1,X_2,X_3)$ as a vector-valued matrix 
$X=\sum_{m=1..3}{X_m\otimes e_m}$, where $(e_m)_{m=1..3}$ is the 
canonical basis of $\C^3$.  Let $R\approx \C^N$ and $Q\approx \C^3$ be the 
vector spaces carrying the regular and defining representations 
of $\Gamma$ respectively. One can think of $Q$ as the covering space of our orbifold and of 
$R$ as the space of Chan-Paton factors. 
Since each of the matrices $X_m$ can be identified 
with a linear operator in $Hom(R,R)$, 
we can view $X$ as an element of $Hom(R,Q\otimes R)$. 
To make everything basis-independent, we let $\rho_R$ denote the regular 
representation of $\Gamma$ and $\rho_Q$ denote the defining representation. 
Then the more concrete formulation of the previous section is obtained by 
picking orthonormal bases $e_m$, ($m=1..3$) of $Q$ and 
$e_\gamma$ $(\gamma \in \Gamma)$ of $R$ and identifying the linear operators 
$\rho_Q(\gamma)$, $\rho_R(\gamma)$ with their matrices $D^{(Q)}(\gamma), 
D^{(R)}(\gamma)$ in these bases. 

\subsection{Quiver data}

In this abstract language, 
the projection conditions require $X$ to be $\Gamma$-invariant in the 
following sense:
\be
\label{inv}
X\rho_R(\gamma)=\rho_{Q\otimes R}(\gamma)X~~.
\ee
As before, let $R_\mu$~($\mu=0..r$) be the irreducible representations 
of $\Gamma$ 
(with $R_0$ the trivial representation) and let $n_\mu:=dim R_\mu$. 
One has the standard decomposition $R=\oplus_{\mu=0..r}{V_\mu\otimes R_\mu}$,  
where the $n_\mu$-dimensional vector spaces $V_\mu$ encode the multiplicities 
of $R_\mu$ as factors of $R$.

Consider the decompositions $Q\otimes R_\nu=\oplus_{\lambda=0..r}
{A_{\lambda \nu}\otimes R_\lambda}$ of the tensor products $Q\otimes R_\nu$ 
into irreducible representations of $\Gamma$, and let 
$a_{\lambda \mu}=dim A_{\lambda \mu}$ be the associated multiplicities. 
From Schur's lemma we know that the subspace $Hom(R_\mu,R_\lambda)^\Gamma$ 
of $\Gamma$-invariant linear maps 
from $R_\mu$ to $R_\lambda $ is zero unless $\lambda =\mu$, 
while the invariant maps from $R_\mu$ to $R_\mu$ are the constant multiples 
of the identity map. Combined with the decompositions discussed above, this 
shows the existence of an isomorphism:
\be
\label{isomf}
Hom(R,Q\otimes R)^\Gamma\approx 
\oplus_{\mu,\nu=0..r}{A_{\mu\nu}\otimes Hom(V_\mu,V_\nu)}~~,
\ee
where $Hom(R,Q\otimes R)^\Gamma$ denotes the subspace of $\Gamma$-invariant 
elements of $Hom(R,Q\otimes R)$. 
Modulo this isomorphism, 
we can identify an element $X\in Hom(R,Q\otimes R)$ which satisfies the 
projection conditions with a set of objects
$\phi^{(\nu\mu)}\in A_{\mu\nu}\otimes Hom(V_\mu,V_\nu)$.
Picking orthonormal 
bases $|\nu\mu;\alpha\rangle$($\alpha=1..a_{\mu\nu}$) of the vector spaces 
$A_{\mu\nu}$, we can write, more specifically: 
\be
\label{chgvar}
\phi^{(\nu\mu)}=\sum_{\alpha=1..a_{\mu\nu}}{\phi^{(\nu\mu)}_\alpha
\otimes |\nu\mu;\alpha\rangle \otimes 1_{n_\mu}}~~, 
\ee
with $\phi^{(\nu\mu)}_\alpha\in Hom(V_\mu,V_\nu)$. 
The linear maps $\phi^{(\nu\mu)}_\alpha:V_\mu\rightarrow V_\nu$ are the 
desired unconstrained variables, which we will call {\em quiver data}. 
Their combinatorial structure can be described in the 
language of graph theory as follows. 

The {\em McKay quiver} is the graph with $r+1$ nodes, 
indexed by $\mu=0..r$, which is obtained by drawing 
$a_{\mu\nu}$ arrows (i.e. oriented edges) starting from the node 
$\mu$ and ending at the node 
$\nu$, for each ordered pair of distinct indices $(\mu,\nu)$ satisfying 
$a_{\mu\nu}\neq 0$ (if both $a_{\mu\nu}$ and $a_{\nu\mu}$ are nonzero, then 
there will be $a_{\mu\nu}$ arrows from $\mu$ to $\nu$ and $a_{\nu \mu}$ 
arrows from $\nu$ to $\mu$), and by drawing $a_{\mu\mu}$ loops 
(i.e. edges connecting a node with itself) at the node  
$\mu$ for each $a_{\mu\mu}\neq 0$. Note that the loops of the quiver do not 
carry any  natural orientation. 
By indexing the arrows from $\mu$ to $\nu$ 
by an integer $\alpha=1..a_{\mu\nu}$, one can think of each arrow as 
being a pictorial representation of the quiver datum 
$\phi^{(\nu\mu)}_\alpha:V_\mu\rightarrow V_\nu$ 
(the same applies to the loops). 
The McKay quiver encodes the branching rules for the decompositions of the 
tensor product representations  $Q\otimes R_\nu$ into irreducible 
representations of $\Gamma$. 

\subsection{The projected gauge group and the moment map}

The structure of the surviving gauge group $G_0:=U(R)^\Gamma$ 
can be found more abstractly as follows. The decomposition of the 
regular representation shows that:
\be
\label{group}
U(R)^\Gamma\approx \Pi_{\mu=0..r}U(V_\mu)\approx \Pi_{\mu=0..r}U(n_\mu)~~.
\ee
This isomorphism is realized by sending an $(r+1)-tuple$ of matrices 
$(g_0...g_{n_r})\in \Pi_{\mu=0..r}U(n_\mu)$ into the block-diagonal 
$N\times N$ matrix $U\in U(R)^\Gamma$ which contains $n_\mu$ copies of 
$g_\mu$ for each $\mu=0..r$:
\be
\label{isom}
\left[\begin{array}{cccc}
U_0\otimes 1_{n_0}&0&...&0\\
0&U_1\otimes 1_{n_1}&...&0\\
...&...&...&...\\
0&...&0&U_r\otimes 1_{n_r}\end{array}\right] \leftrightarrow 
\left[\begin{array}{cccc}
U_0 &0&...&0\\
0&U_1 &...&0\\
...&...&...&...\\
0&...&0&U_r \end{array}\right]~~.
\ee 

The action of $U(R)^\Gamma$ on the matrices $X$ translates into the natural 
action of $\Pi_{\mu=0..r}U(V_\mu)$ on the quiver data:
\be
\phi^{(\nu\mu)}_\alpha\rightarrow g_\nu\phi^{(\nu\mu)}_\alpha (g_\mu)^{-1}~~,
\ee
for all $g=(g_0...g_r)\in \Pi_{\mu=0..r}U(V_\mu)$. Since 
the diagonal $U(1)$ subgroup of $G_0$ acts trivially on all such data, 
the group which acts effectively on their space is:
\be
G=\Pi_{\mu=0..r}U(V_\mu)/U(1)_{diag}~~,
\ee
which we will call {\em the effective gauge group}.

The moment map for the action of $G$ on the quiver data
is given by general theory\footnote{Here we consider the natural hermitian 
product $<\Phi,\Psi>=\sum_{\scriptsize\begin{array}{c}\mu,\nu=0..r\\
\alpha=1..a_{\mu\nu}\end{array}}
{(\phi^{(\nu\mu)}_\alpha)^+\psi^{(\nu\mu)}_\alpha}$
on the vector space ${\cal Q}$ of all quiver data, 
which is preserved by the action of $G$. It 
is well-known that the moment map $M:{\cal Q}\rightarrow {\bf g}$ 
for such an action is given by the relation 
$Tr(M(\Phi)\theta)=<\Phi, \theta \Psi>$, where 
$\theta \in {\bf g}$ is arbitrary and $\Phi,\Psi$ is any pair of quiver data
(here ${\bf g}$ is the Lie algebra of $G$). 
Using the presentation (\ref{group}) of $G$ gives a realization of 
${\bf g}$ as the `traceless' subalgebra of 
$\oplus_{\mu=0..r}{u(n_\mu)}$, so we can write 
$\theta =\oplus_{\mu=0..r}{\theta_\mu}$, with $\theta_\mu\in u(n_\mu)$ 
and $M(\Phi)=\oplus_{\mu=0..r}{M_\mu(\Phi)}$, where 
$M_\mu$ are maps from ${\cal Q}$ to $u(n_\mu)$. 
Using arbitrariness of $\theta_\mu$, this immediately leads to the expression 
(\ref{M}). Note that the original moment map $\rho$ is associated to the 
hermitian product $<X,Y>=Tr(X_m^+Y_m)$, which differs from the product we 
used above. Indeed, if $\Phi=(\phi^{(\nu\mu)}_\alpha),
\Psi:=(\psi^{(\nu\mu)}_\alpha)$ are the quiver data associated to $X,Y$, then 
$Tr(X_m^+Y_m)=\sum_{\scriptsize\begin{array}{c}\mu,\nu=0..r\\
\alpha=1..a_{\mu\nu}\end{array}}
{n_\mu(\phi^{(\nu\mu)}_\alpha)^+\psi^{(\nu\mu)}_\alpha}$. The factors of 
$n_\mu$ occur because of the presence of the identity map $1_{n_\mu}$ 
in the decomposition (\ref{chgvar}).} as:
\be
\label{Mmu}
M=\oplus_{\mu=0..r}M_\mu~~,
\ee
where:
\be
\label{M}
M_\mu=\sum_{\nu,~a_{\nu\mu}\neq 0}{\sum_{\alpha=1..a_{\nu\mu}}
{\phi^{(\mu\nu)}_{\alpha}(\phi^{(\mu\nu)}_\alpha)^+}}
-\sum_{\nu,~a_{\mu\nu}\neq 0}
{\sum_{\alpha=1..a_{\mu\nu}}
{(\phi^{(\nu\mu)}_\alpha)^+\phi^{(\nu\mu)}_\alpha}} \in u(n_\mu)~~.
\ee
The map $M_\mu$ is related to the original moment map $\rho$ of equation 
(\ref{moment1}) through the isomorphism  
of Lie algebras induced by (\ref{isom}). In fact, it is not hard to see
\footnote{This follows easily by noting that $\rho$ has the form:
$\rho(X)=\oplus_{\mu=0..r}{\rho_\mu(X)}^{\oplus n_\mu}$, with 
$\rho_\mu$ a map from ${\cal Q}$ to $u(n_\mu)$. The relation between 
$\rho$ and $M$ is given by the condition 
$Tr(M(\Phi_X)\theta )=Tr(\rho(X)q(\theta))$ for all 
$\theta=\oplus_{\mu=0..r}{\theta_\mu}\in \oplus_{\mu=0..r}{u({n_\mu})}$, 
where 
$q(\theta)=\oplus_{\mu=0..r}{(\theta_\mu)^{\oplus n_\mu}}$ is 
the isomorphism induced by (\ref{isom}) and $\Phi_X=(\phi^{(\nu\mu)}_\alpha)$
is the set of quiver data associated to $X$. 
The factors $n_\mu$ in the relation between $\rho$ and 
$M$ appear when one evaluates the traces, which gives $\rho_\mu(X)=
\frac{1}{n_\mu}M_\mu(\Phi_X)$.}
that:
\be
\label{mom_relation}
\rho=\oplus_{\mu=0..r}{\frac{1}{n_\mu}(M_\mu)^{\oplus n_\mu}}~~.
\ee
The `quiver moment map' (\ref{M}) has a simple graphical interpretation 
(see Figure 1), obtained by noting that the two terms of (\ref{M}) 
correspond to summing over all arrows leaving, respectively entering a given 
node $\mu$ (if there is a loop at that node, then it is considederd in both 
sums, i.e. it is viewed as both leaving and entering the node).

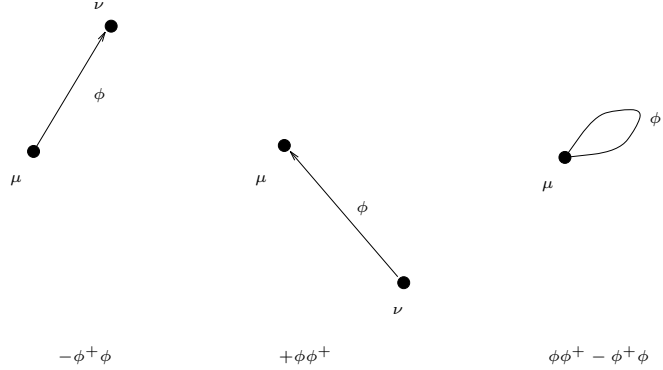
\begin{figure}[htpb]
\begin{center}
\input{moment.pstex_t}
\caption{\footnotesize Three types of quiver data $\phi$ related to a node 
and their contribution to the quiver moment map.}
\label{Figure 1.}
\end{center}
\end{figure}

The central levels of $M$ correspond to $(r+1)$-tuples 
$(\xi_0...\xi_r)\in \R^{r+1}$ satisfying $\sum_{\mu=0..r}{n_\mu\xi_\mu}=0$, 
which amounts to writing the original matrix $\zeta$ of equation (\ref{zeta}) 
in the form 
$\zeta=diag(\frac{\xi_0}{n_0} 1_{n^2_0}...\frac{\xi_r}{n_r} 1_{n^2_r})$.
The moment map equations become $M={\hat \xi}$, 
where ${\hat \xi}=\oplus_{\mu=0..r}{\xi_\mu 1_{n_\mu}}$, i.e.:
\be
M_\mu=\xi_\mu 1_{n_\mu}~~(\mu=0..r)~~.
\ee

It is convenient to identify the group $G$ with:
\be
\label{G_final}
G\approx S[\Pi_{\mu=0..r}U(n_\mu)]/K~~,
\ee
where $S[\Pi_{\mu=0..r}U(n_\mu)]$ is the group of $(r+1)$-tuples 
$(g_0...g_r)\in \Pi_{\mu=0..r}U(n_\mu)$ satisfying $\Pi_{\mu=0..r}det(g_\mu)=1$
and $K\approx \Z_n$~($n:=\sum_{\mu=0..r}{n_\mu}$) is the finite central 
subgroup given by:
\be
K=\{(\eta 1_{n_0}....\eta 1_{n_r})|\eta^n=1\}~~.
\ee
From now on, we will always use this presentation of $G$. 
Due to the tracelessness condition $\sum_{\mu=0..r}{n_\mu \xi_\mu}=0$, 
the values of the Fayet-Iliopoulos parameters can be described by the 
vector $\xi=(\xi_1...\xi_r) \in \R^r$. We will denote 
the space of such vectors by $\R^r(\xi)$.

\subsection{Quiver Feynman rules for the superpotential}

The parametrization of the variety of commuting matrices in terms of  
quiver data can be obtained by first expressing the superpotential in the 
variables $\phi^{(\nu\mu)}_\alpha$.  Then ${\cal Z}$ can be obtained as the 
critical set of $W$.

In order to express the superpotential in terms of $\phi^{(\nu\mu)}_\alpha$, 
we need to know the  explicit form of the equivalence (\ref{isomf}). 
This can be obtained as follows. 
For each irreducible representation $R_\mu$ of $\Gamma$, pick a specific 
realization of $R_\mu$ via unitary matrices $D^{(\mu)}(\gamma)\in U(n_\mu)$.
Given a hermitian vector space $S$ carrying a unitary 
representation $\rho$ of $\Gamma$ which is equivalent to $R_\mu$,
an orthonormal basis $(e^{(\mu)}_i)_{i=1..n_\mu}$ of $S$ 
will be called {\em fiducial} if $\rho(\gamma)e^{(\mu)}_i=
D^{(\mu)}_{ji}(\gamma)e^{(\mu)}_j$, 
for all $\gamma \in \Gamma$. By Schur's lemma, such a basis is determined up 
to a global phase factor. If the representation $\rho$ carried by $S$ is 
equivalent to $aR_\mu=R_\mu^{\oplus a}$, then an orthonormal basis 
$(e^{(\mu,\alpha)}_i)_{\alpha= 1..a; i=1..n_\mu}$ of $S$ is called fiducial if 
$\rho(\gamma)e^{(\mu,\alpha)}_i=D^{(\mu)}_{ji}(\gamma)e^{(\mu,\alpha)}_j$ for 
all $\alpha=1..a$ and all $\gamma$ in $\Gamma$. A fiducial 
basis in this generalized sense is determined 
up to a transformation of the form 
$e^{(\mu,\alpha)}_i\rightarrow U_{\alpha,\beta}e^{(\mu,\beta)}_i$, with 
$U=(U_{\alpha\beta})_{\alpha,\beta=1..a}$ a unitary $a \times a$ matrix. 
(The notion of fiducial bases is necessary in order to fix our conventions on 
the Clebsch-Gordan coefficients below.) 

Consider the decomposition:
\be
Q\otimes R_\nu=\oplus_{\mu=0..r}{A_{\mu \nu}\otimes R_\mu}~~.
\ee
Picking fiducial bases $e^{(Q)}_j$, $e^{(\nu)}_k$ ($j=1..n_Q,k=1..n_\nu$) 
of $Q$, respectively $R_\nu$ and $e^{(\mu,\alpha)}_i$ 
($i=1..n_\mu,\alpha=1..a_{\mu\nu}$) of $A_{\mu \nu}\otimes R_\mu$, we can 
write:
\be
e^{(\mu,\alpha)}_i=\sum_{\scriptsize 
\begin{array}{c}j=1..n_Q\\k=1..n_\nu\end{array}}{C^{\mu i,\alpha}_{Qj,\nu k}
e^{(Q)}_j\otimes e^{(\nu)}_k}~~, 
\ee
where $C^{\mu i,\alpha}_{Qj,\nu k}$ are the Clebsch-Gordan 
coefficients relating these bases. Then it is easy to see that any 
$X$ satisfying (\ref{inv}) has the form:
\be
X=\oplus_{\mu,\nu=0..r}{\sum_{\alpha=1..a_{\mu\nu}}
{\phi^{(\nu\mu)}_\alpha\sum_{i=1..n_\mu}{|e_i^{(\mu,\alpha)}
\rangle\langle e_i^{(\mu)}|}}}~~,
\ee
while its components are given by:
\be
\label{Xm}
X_m:=\oplus_{\mu,\nu=0..r}{\sum_{\scriptsize \begin{array}{c}
i=1..n_\mu\\j=1..n_\nu\end{array}}{
\sum_{\alpha=1..a_{\mu\nu}}{C^{\mu i,\alpha}_{3m,\nu j}\phi^{(\nu \mu)}_\alpha
|e_j^{(\nu)}\rangle \langle e_i^{(\mu)}|}}}~~.
\ee
This relation allows us to rewrite the superpotential (\ref{superpot}) as:
\be
\label{Wquiver}
W=-\sum_{\scriptsize
\begin{array}{c} \mu,\nu,\lambda=0..r \\
(a_{\mu\nu}a_{\nu\lambda}a_{\lambda\mu}\neq 0)\end{array}}{
\sum_{\scriptsize 
\begin{array}{c}i=1..n_\mu \\j=1..n_\nu \\k=1..n_\lambda\end{array}}
{\sum_{\scriptsize 
\begin{array}{c}\alpha=1..a_{\mu\nu}\\\beta=1..a_{\nu\lambda}\\
\gamma=1..a_{\lambda\mu}\end{array}}
{\epsilon_{mnl}C^{\mu i,\alpha}_{3m,\nu j}C^{\nu j,\beta}_{3n,\lambda k}
C^{\lambda k,\gamma}_{3l,\mu i} 
Tr[\phi^{(\mu\lambda)}_\gamma\phi^{(\lambda\nu)}_\beta\phi^{(\nu\mu)}_\alpha]}}}~~.
\ee
This expression admits a graphic description which we now explain. 
Given a quiver, one can construct its {\em reduction}, which is the 
quiver obtained by keeping only one edge  
out of each set of edges connecting any two given nodes 
(the two nodes under consideration need not be distinct, so that reduction is 
also applied to any loops which may be present in the quiver). 
This amounts to leaving the nodes $\mu$ unchanged 
and replacing $a_{\mu\nu}$ by ${\tilde a}_{\mu\nu}=
\{\begin{array}{c}  1, \mbox{~if~}a_{\mu\nu}\neq 0\\
			0, \mbox{~if~}a_{\mu\nu}=0
\end{array}$, for all $\mu,\nu=0..r$. 
The reduced quiver contains at most one edge between any two nodes. 

A {\em triple circuit} of a quiver is an ordered triplet of edges 
\footnote{ Since our use of the word circuit may be slightly unfamiliar, 
let us give a formal definition of the objects involved. 
In precise set-theoretic language, a quiver is a 
quadruplet $(S,V,h,t)$ 
where $S$ is a finite set whose elements are called {\em edges},  
$V$ is a finite set whose elements are called {\em nodes} and $h,t$ 
(also denoted by $head, tail$) are maps from $S$ to $V$. 
If $s \in S$ is an edge, then the nodes $h(s),t(s)\in V$ are called the 
{\em head}, respectively the {\em tail} of $s$. An edge is called 
an {\em arrow} if $t(s)\neq h(s)$ 
(one represents this graphically by drawing an 
arrow from $t(s)$ to $h(s)$). Otherwise, $s$ is called a {\em loop} (the 
graphical representation of a loop does not carry an orientation).  
A {\em marked circuit} is a finite ordered set 
$(s_0...s_k)$ of edges such that 
$h(s_{j-1})=t(s_j)$ for all $j=1..k$ and $h(s_k)=t(s_0)$. Two marked circuits 
$(s_0...s_k)$ and $(s'_0...s'_k)$
are called {\em equivalent} if there exists a {\em cyclic} permutation 
$\sigma$ of the set $\{0..k\}$ such that $s_{\sigma(i)}=s'_i$ for all $i=0..k$
(this defines an equivalence relation). A {\em circuit} is an equivalence 
class of marked circuits modulo this equivalence relation. Intuitively, 
a marking of a circuit is given by chosing a `starting point' for traversing 
the circuit, while the circuit itself is obtained by `forgetting' the 
marking.}
$(f,g,h)$ (not necessarily distinct and considered only up to a circular 
permutation) such that 
$tail(g)=head(f)$, $tail(h)=head(g)$, $tail(f)=head(h)$ 
(For a loop $l$ we define $head(l)=tail(l)$ to be the associated node.)  
The {\em nodes} of the circuit are the (not necessarily distinct) points 
$tail(g)=head(f)$, $tail(h)=head(g)$, $tail(f)=head(h)$ which the circuit 
meets. 

A set of  `Feynman rules' for computing the superpotential $W$ 
can now be formulated in terms of the reduced quiver: 

(1) To obtain the total superpotential, one must add the contribution of all 
different triple circuits of the reduced quiver.

(2)For any  circuit of the reduced quiver,
pick any of the nodes of the circuit and let $\mu$ denote its index. 
Then follow the circuit in the sense given by its orientation and let 
$\nu,\lambda$ be the indices of the first and second nodes thus encountered
(note that $\mu,\nu,\lambda$ need not be distinct). Then 
there is a contribution to the superpotential given by (see Figure 2):
\be
\label{feynman}
-w\sum_{\scriptsize 
\begin{array}{c}i=1..n_\mu \\j=1..n_\nu \\k=1..n_\lambda\end{array}}{
\sum_{\scriptsize\begin{array}{c}
\alpha=1..a_{\mu\nu}\\\beta=1..a_{\nu\lambda}\\\gamma=1..a_{\lambda\mu}
\end{array}}
{\epsilon_{mnl}C^{\mu i,\alpha}_{3m,\nu j}C^{\nu j,\beta}_{3n,\lambda k}
C^{\lambda k,\gamma}_{3l,\mu i}
Tr[\phi^{(\mu\lambda)}_\gamma\phi^{(\lambda\nu)}_\beta
\phi^{(\nu\mu)}_\alpha]}}~~,
\ee 
where $w$ is a multiplicity equal to:

(a)$w$=3, unless all 3 nodes coincide

(b)$w$=1, if all 3 nodes coincide. 

\noindent (This multiplicity is a consequence of the invariance of the sum in 
(\ref{feynman}) under cyclic permutations of the triplet $(\mu,\nu,\lambda)$). 
The types of circuits with 3 nodes which can occur in a reduced quiver 
are drawn in Figure 3. The third of these consists of a loop which is 
traversed 3 times and is not orientable. 

\

\iffigs
$$\vbox{
\hskip 1.0 in \hbox{\epsfxsize=7cm\epsfbox{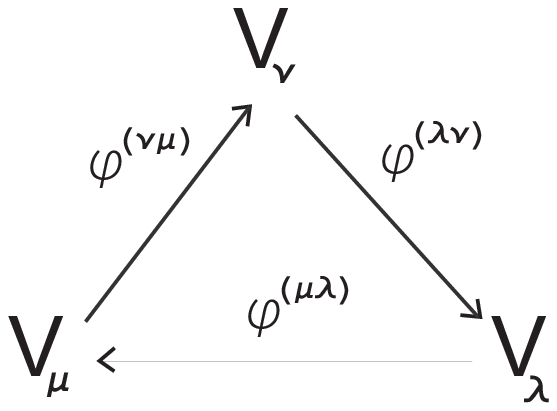}}

\hskip 0.5 in \hbox{Figure 2. {\footnotesize  The graphic structure of a 
contribution to the superpotential.}}}$$
\fi

\

\

\iffigs
$$\vbox{\hskip 0.8 in \hbox{\epsfxsize=9cm\epsfbox{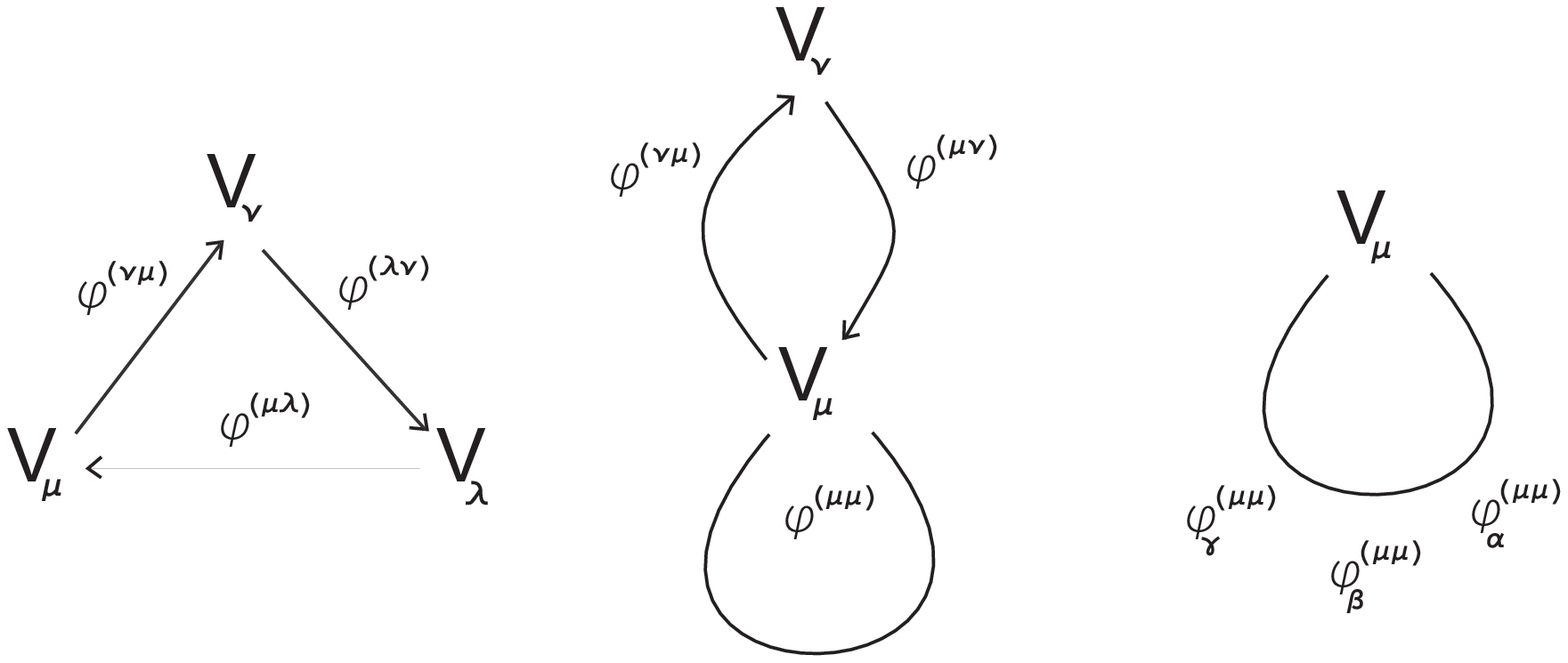}}

\hskip 0.6 in \hbox{Figure 3. {\footnotesize Types of oriented circuits 
allowed in a reduced quiver.}} 

\hskip 0.6 in \hbox{{\footnotesize 
The last type of circuit consists of a loop traversed 3 times.}}}$$
\fi
\

\subsection{Comparison with the case when $\Gamma$ is abelian}

Let us pause for a moment to discuss how the case of an abelian 
orbifold group fits into the above formalism. If $\Gamma$ is abelian, 
then all of its irreducible representations are one-dimensional. The number 
$r+1$ of irreducible representations is equal to the number $N$ of elements of 
$\Gamma$ and $n_\mu=1$ for all $\mu=0..N-1$. The spaces $V_\mu$ sitting at 
the nodes of the quiver are therefore one-dimensional. Chosing a nonzero 
vector $u_\mu \in V_\mu$ for each $\mu$ allows us to identify 
$V_\mu\approx \C$.
The quiver data 
$\phi^{(\nu\mu)}_\alpha:V_\mu\rightarrow V_\nu$ can be  identified with the 
complex numbers $x^{(\nu\mu)}_\alpha$ given by: 
$\phi^{(\nu\mu)}_\alpha(u_\mu)=x^{(\nu\mu)}_\alpha u_\nu$. In fact, 
$x^{(\nu\mu)}_\alpha$ are nothing else than those  
components of the matrices $X_m$ which survive the projection conditions. 
Since all $n_\mu=1$, the projected gauge 
group is abelian, and it coincides with the compact torus $G_0=U(1)^N$, 
while the effective gauge group is given by 
$G=U(1)^N/U(1)_{diag}\approx U(1)^{N-1}$. 
Moreover, equation 
(\ref{mom_relation}) shows that the original 
moment map $\rho$  coincides with the quiver moment map $M$. 

The moduli space 
is given by a K\"{a}hler reduction of the variety of commuting matrices via 
the action of $U(1)^{N-1}$, which is equivalent 
(by results of \cite{Audin}) with a holomorphic quotient of ${\cal Z}$ by 
the complex torus $(\C^*)^{N-1}$. The important fact which simplifies the 
analysis in this case (and which was pointed out in 
\cite{branes3, Infirri}) 
is that the variety of commuting matrices is a toric variety. 
(This can be established, for example, 
by writing down the commutation relations $[X_m,X_n]=0$ in terms 
of those components of $X_m$ which survive the projection conditions and 
noticing that the resulting F-flatness constraints are given by monomial 
relations, which assures that ${\cal Z}$ is an affine toric variety. 
A similar argument can be made at the level of the quiver 
data $x^{(\nu\mu)}_\alpha$.) Using 
the alternative presentation \cite{Cox}  of ${\cal Z}$  
as a holomorphic quotient of a space $\C^d$ by a complex torus $(\C^*)^k$ 
allows one to present the moduli space ${\cal M}_\xi$ itself as a toric 
variety. The variation of ${\cal M}_\xi$ with $\xi$ can then be studied by the 
well-established methods of toric geometry \cite{toric}. 

By contrast with the above, nonabelian groups $\Gamma$ lead to varieties 
${\cal Z}$ which are in general not toric. This is due to the fact that, 
in the general case, some of the irreducible representations of $\Gamma$ are 
not one-dimensional (some $n_\mu$ is different from $1$), so that the projected
fields $X_m$ have a complicated block structure. This leads to 
{\em polynomial} (as opposed to monomial) constraints among the surviving 
components of $X_m$, when one imposes the commutation conditions $[X_m,X_n]=0$ 
which define the variety of commuting matrices. Alternatively, this can be 
seen at the level of quiver data, if one notices that the traces 
$Tr[\phi^{(\mu\lambda)}_\gamma\phi^{(\lambda\nu)}_\beta
\phi^{(\nu\mu)}_\alpha]$ appearing in the quiver Feynman rules 
(\ref{feynman}) will produce polynomials in the components of the maps 
$\phi$ if one of the vector spaces $V_\mu,~V_\nu,~V_\lambda$ has dimension 
greater than one. 

On the other hand, the effective gauge group $G$ of equation (\ref{G_final})
is in general nonabelian, which means that, even if one has obtained 
an explicit description of ${\cal Z}$, the moduli space is given by a 
K\"{a}hler reduction of ${\cal Z}$ via a {\em nonabelian} group. 
While one can still use results of \cite{Kirwan} to reduce this to a 
geometric invariant theory quotient of ${\cal Z}$ by the complexified gauge 
group $G^\C$, the computation of such a quotient is a highly nontrivial 
problem in algorithmic invariant theory \cite{GITconstructive}. 
In the next section we will investigate the variation of the moduli space 
as a function of the Fayet-Iliopoulos parameters 
by using methods of symplectic and algebraic geometry.

\section{The variation of the moduli space}

The study of the moduli space of D-branes at nonabelian quotient singularities 
of threefolds is complicated by the lack of efficient computational 
methods for treating the resulting symplectic quotient problem. 
This is in contrast with the case of abelian groups $\Gamma$,
where the problem can reduced to toric 
geometry, as was schetched above following the detailed discussion of 
\cite{branes3,Infirri}. 
The nonabelian case is considerably 
more complicated, since in this situation the moduli space ${\cal M}_\xi$ is 
not a toric variety, so one cannot resort to the powerful machinery of 
\cite{toric} to reduce the problem to convex geometry. (Apparently, then, 
the only tools available in such a situation are the general methods of 
geometric invariant theory \cite{GIT}, which lead to important qualitative 
results but to hard computational problems. The determination of the moduli 
space could be achieved, at least in principle, by a computation of 
`relative invariants' in the spirit of classical invariant theory
\cite{GITconstructive}. Unfortunately, this problem is extremely involved 
in practice and severely hampers the hope for progress). 

However, one can extract considerable information about 
the {\em variation} of the moduli space without performing explicit 
calculations of invariants, and, rather suprisingly, one can reduce some basic 
questions about this variation to problems which can be handled by toric 
methods. While the methods we discuss below are 
not powerful enough to determine the moduli space explicitly, or 
to compute its geometric properties, they 
suffice to locate the possible topology-changing transitions in the space of 
Fayet-Iliopoulos parameters.

In order to explain this, let us first outline the general picture 
for the dependence of ${\cal M}_\xi$ on $\xi$, 
which follows from the results of 
\cite{GITflips1,GITflips2,Kirwan}. 
The allowed levels 
$\xi \in \R^r$ can be divided into `noncritical' and `critical' 
values. The critical values occur on {\em walls} of codimension at least 
one, which separate the space of Fayet-Iliopoulos parameters into disjoint 
chambers. In our case, these chambers will be cones adjoining each other along 
common faces, which determine the walls. The points lying in the interior of 
each chamber are the noncritical values of $\xi$. The main results of 
\cite{GITflips1,GITflips2,Kirwan} state that the quotient ${\cal M}_\xi$ 
depends only on the chamber to which $\xi$ belongs, and can undergo a 
topology-changing transition as $\xi$ crosses a wall. 
Furthermore, the arguments of \cite{GITflips1,GITflips2} 
imply that, if ${\cal M}_\xi$ is smooth for $\xi$ inside any of the chambers, 
then this transition is given by a {\em flip}
\footnote{A flip is a generalization of 
a flop, as we explain in more detail in the next subsection.}. If, moreover, 
the desingularizations ${\cal M}_\xi\rightarrow \C^3/\Gamma $ are crepant 
(i.e. if ${\cal M}_\xi$ is a Calabi-Yau manifold for generic values of $\xi$),
then one can invoke results of \cite{flops} in order to deduce 
that such transitions are in fact {\em flops}. Hence the 
situation is similar to the more familiar cases of topology change 
\cite{topchange} which are realized in the moduli space of two-dimensional 
conformal field theories. 

In view of the general expectations above, the most important information to 
extract is the {\em phase structure} of the space of Fayet-Iliopoulos 
parameters, i.e. the location of the system of walls and chambers in the 
space $\R^r(\xi)$.  As we discuss below, and illustrate explicitly with two 
examples in the next section, this information can be obtained by methods of 
toric geometry, thus bypassing the computational difficulties involved in 
determining the precise structure of the moduli space.  
This follows from the observation that one 
can perform the K\"{a}hler quotient 
${\cal M}_\xi=({\cal Z}\cap M^{-1}(\xi))/G$ in two steps, if one 
writes $G=(T\times H)$ with $T$ a central Lie subgroup of $G$ and 
$H\approx G/T$
(where we neglect an unimportant finite central subgroup). 
Then the moduli space 
${\cal M_\xi}$ turns out to be isomorphic with the zero level 
K\"{a}hler quotient 
of $P_\xi:=({\cal Z}\cap M_T^{-1}(\xi))/T$ by the induced action of $H$. 
(Here $M_T$ is the projection of the moment map $M$ onto the center 
of the Lie algebra of $G$).  The salient point is that only the first step of 
this double reduction depends on $\xi$, which suggests that the  
$\xi$-dependence of the moduli space is essentially encoded 
by the quotient $P_\xi=({\cal Z}\cap M_T^{-1}(\xi))/T$. 
This can be studied by methods 
of toric geometry, since $T$ will be a compact torus $U(1)^r$ (up to some 
finite identifications). Indeed, since ${\cal Z}$ is an affine variety inside 
of ${\cal Q}$, this quotient presents $P_\xi$ as a 
subvariety (given by homogeneous polynomial equations) of the space 
$E_\xi=\{\Phi \in {\cal Q}|M_T(\phi)=\xi\}/T$, which is a toric variety. 
The wall structure of $\R^r(\xi)$ associated to the quotient $E$ can be 
easily determined by toric methods, and the walls associated to $P_\xi$ (and 
thus to ${\cal M}_\xi$) turn out to form a `coarsening' of the set of walls 
associated to $E_\xi$.

The plan of this section is as follows. In the first subsection, we discuss 
the basic features of the variation of the moduli space which follow from 
general results in symplectic geometry and geometric invariant theory. 
(The reader unfamiliar with algebraic geometry and invariant theory 
can safely skip this section, which is not strictly needed for understanding 
the rest of the paper.)
Subsection 3.2 outlines the double quotient 
procedure for the computation of ${\cal M}_\xi$, while in subsection 3.3 
we argue that the variation of the moduli space with the Fayet-Iliopoulos 
parameters $\xi$ is essentially encoded by the toric part of the quotient. 
Finally, subsections 3.4 and 3.5 discuss the toric part of the quotient in 
detail. 

\subsection{Flips and flops}

In order to substantiate the above discussion, it is most convenient to 
translate the problem to one in algebraic geometry, since this will allow 
us to apply the results of \cite{GITflips1,GITflips2}. 

For this, first note that ${\cal Z}$ is an affine algebraic variety 
in the ambient vector space ${\cal Q}$, since 
it is given by a finite set of polynomial equations among the quiver 
variables $\phi^{(\nu\mu)}_\alpha \in {\cal Q}$. 
Moreover, the universal cover ${\tilde G}=S[\Pi_{\mu=0..r}{U(n_\mu)}]$
of the effective gauge group  $G$ 
is a product of a semisimple classical group and 
a torus. It this case, it is well-known that the complexification 
${\tilde G}_\C=S[\Pi_{\mu=0..r}{SL(n_\mu)}]$ of ${\tilde G}$ is a 
{\em reductive}
\footnote{A complex algebraic group $U$ is {\em reductive} if any 
finite-dimensional rational representation of $U$ is completely reducible. 
A rational representation of $U$ is simply a matrix representation  
such that all entries of the representation matrix $A(u)$ are rational 
functions of the group element $u$.}
algebraic group, whose induced action on ${\cal Z}$ is rational. 
The moduli space ${\cal M}_\xi$ will be a (quasiprojective) algebraic variety 
only if the Kahler form induced by the Kahler quotient procedure is integral.
It is not hard to see that this will be the case if one restricts to integral 
levels of the moment map, i.e. Fayet-Iliopoulos terms  
$\xi=\sum_{\mu=0..r}{\xi_\mu 1_{n_\mu}}$ whose components $\xi_\mu$ are 
integers.
In this case, one can apply results of \cite{Kirwan} in order to deduce 
that the K\"{a}hler reduction ${\cal M}_\xi$ of ${\cal Z}$ by $G$ at level 
$\xi$ is isomophic (as a complex variety) with the geometric invariant theory 
quotient ${\cal Z}//_{\chi_\xi}G_\C$
of ${\cal Z}$ by $G_\C$ linearized
\footnote{In our case, a linearization of the action of $G_\C$ on $X$ 
is a lift to an action on 
the total space of the trivial holomorphic line bundle ${\cal O}_X$, which 
agrees with the action on $X$. This is the same as a choice of a rational 
character of $G_\C$.}
by the rational character associated to $\xi$:
\be
\chi_\xi(g_0..g_r)=(detg_0)^{\xi_0}...(det g_r)^{\xi_r}~~(g_\mu \in SL(n_\mu)).
\ee
In fact, one can allow $\xi$ to be rational \cite{GITflips1,GITflips2} 
by considering `fractional linearizations', 
so the integrality restriction on $\xi$ is of little consequence. 

Once the problem has been reduced to one in algebraic geometry, one can 
establish the following results. First, the space of moment map levels 
is divided into a finite number of {\em chambers}, which adjoin along 
a set of walls. These chambers and walls can be shown to be 
{\em integral convex polytopes} with respect to the sublattice determined by 
the integral moment map levels. 
The K\"{a}hler reduction ${\cal M}_\xi$ is unchanged
(up to an isomorphism of {\em complex} varieties) as long as $\xi$ varies 
inside of a given chamber, and it suffers a transition known 
as a {\em flip} when $\xi$ crosses a generic wall. 

Intuitively, a flip is a birational transformation during which a subvariety 
of ${\cal M}_\xi$ of codimension al least $2$ shrinks to a point and is then 
blown up to another such subvariety. More technically, a flip is defined as 
a birational transformation 
$f:X_+\rightarrow X_{-}$ between two 
algebraic varieties which can be decomposed into two birational maps 
$X_+\rightarrow X_0$ and (the inverse of) $X_{-}\rightarrow X_0$, such that 
$X_\pm\rightarrow X_0$ are {\em small contractions}. 
A small contraction $X_+\rightarrow X_0$ is a proper birational morphism 
from $X_+$ to $X_0$ whose exceptional set (the subset of $X_+$ over which 
the map is not one to one) has codimension at least two in $X_+$.  
For a flip, one also makes the technical requirement that there exist 
a divisor $D$ on $X_+$ such that ${\cal O}(-D)$ and the inverse of its 
pushforward $f_*({\cal O}(D))$ on $X_{-}$ are both relatively ample over 
$X_0$\footnote{The reader familiar with Mori theory should note that we do 
not require that $D$ is the canonical divisor; therefore, a flip in the above 
sense is apriori more general than a Mori flip.}. 

According to the discussion above, the moduli space ${\cal M}_\xi$ of our 
D-brane effective field theories will undergo a flip when $\xi$ crosses a 
generic wall. On the other hand, general results of \cite{flops} assure us 
that two {\em crepant} partial resolutions of a Calabi-Yau threefold 
singularity can always be related by a sequence of flops. This allows us to 
conclude that ${\cal M}_\xi$ will in general undergo a flop when it 
crosses a wall in the space of Fayet-Iliopoulos parameters.

\subsection{The double K\"{a}hler quotient}

Consider the K\"{a}hler reduction of ${\cal Z}$ at level $\xi$ 
modulo $T$:
\be
\label{first}
P_\xi:=\{\Phi \in {\cal Z}|M_T(\Phi)=\xi\}/T~~.
\ee
Since $T$ is a central subgroup of $G$, one has an induced action of 
$H:=G/T$ on the quotient $P_\xi$. This action turns out to be 
K\"{a}hlerian with respect to the K\"{a}hler form induced on $P_\xi$ 
by the K\"{a}hler reduction procedure. Denoting the associated 
moment map by $M_H$, we can consider 
the K\"{a}hler quotient of $P_\xi$ by the action of $H$ 
{\em at level zero}:
\be
{\cal N}_\xi:=\{x\in P_\xi|M_H(x)=0\}/H~~.
\ee
Then we claim that:
\be
{\cal M}_\xi\approx {\cal N}_\xi~~,
\ee
as K\"{a}hler manifolds. An outline of the arguments involved in establishing 
this fact can be found in Appendix 1.

\subsection{Identifying the walls of the extended K\"{a}hler cone}

\subsubsection{Variation of the toric part of the quotient}

The first stage (\ref{first}) in our procedure is a quotient 
of ${\cal Z}$ by a hamiltonian torus action
\footnote{The action is hamiltonian with respect to the symplectic form 
given by the imaginary part of the K\"{a}hler form.}. 
Such a situation has been studied 
extensively in the literature and can be approached either from a symplectic 
\cite{Audin} or a toric \cite{toric} perspective. In fact, the quotient 
(\ref{first}) realizes 
$P_\xi$ as a submanifold of the ambient space 
$E_\xi=\{\Phi \in {\cal Q}|M(\Phi)=\xi\}/T$. As a complex 
manifold, this ambient space is a toric variety of the form 
$(\C^k-Z_\xi)/(\C^*)^r$ (where $k=dim{\cal Q}$) and $Z_\xi$ a subset  
of $\C^n$ (a union of intersections of coordinate hyperplanes), called the 
`exceptional set'.
It is well-known that the chambers of $\R^r(\xi)$ for $E_\xi$  
consist of a finite set of polyhedral cones $\sigma_i$, and that $Z_\xi$ 
is constant when $\xi$ varies inside of a given chamber.
Hence chosing $\xi$ in the interior of the cone $\sigma_i$ gives a toric 
variety 
$E_\xi=E_i=({\cal Q}-Z_i)/(\C^*)^r\approx ({\cal Q}-Z_i)/(\C^*)^r$, 
which  depends only on the cone $\sigma_i$. The spaces $E_i$ are related 
by {\em toric flips} which occur when $\xi$ crosses one of the walls, a fact 
which underlies the presence of topology 
changing-transitions in the moduli space of $(2,2)$ conformal field theories
considered in \cite{topchange}. Indeed, when $\xi$ crosses a wall, 
the $(\C^*)^r$ quotients of the exceptional set $Z$ on the two sides of the 
wall will flip. More precisely, if $E_+=({\cal Q}-Z_+)/(C^*)^r$ and 
$E_-=({\cal Q}-Z_{-})/(C^*)^r$ are 
the toric varieties obtained on the two sides of a generic wall, then 
$E_+$ contains the subvariety $(Z_{-}-Z_+\cap Z_{-})/(C^*)^r$, while 
$E_{-}$ contains the subvariety $(Z_{+}-Z_+\cap Z_{-})/(C^*)^r$, and these two 
subvarieties are exchanged during the flip. In fact, the walls for $E_\xi$ can
be identified by methods of toric geometry or with the help of standard 
results of (\cite{Audin}) which characterize the regular levels of the 
moment map for torus actions. 

The underlying complex variety of $P_\xi$ is given by the quotient 
$({\cal Z}-(Z_i\cap {\cal Z}))/(\C^*)^r$, and will undergo similar 
transformations as $\xi$ crosses a wall, since $Z$ changes for those 
values of $\xi$. In fact, the set of walls  for $P_\xi$ will in general be a 
`coarsening'  of the set of walls for $E_\xi$, 
since it is  possible that 
the intersection of ${\cal Z}$ with $Z$ stays unchanged 
even though $Z$ changes during a flip of $E$. 
Hence the only remaining issue 
is to check whether some of the ambient walls are 
`projected out' or identified with other walls when restricting to 
${\cal Z}$, and this can be achieved in 
each case by a study of the intersection of the 
variety of commuting matrices with the various exceptional sets $Z_i$.
\footnote{In the toric case of \cite{branes3,Infirri}, such a projection 
often occurs and can be formalized elegantly in the quiver 
language \cite{Infirri}. 
The situation is more involved in our case, and we were not able 
to find a comparably powerful description.}

\subsubsection{Variation of the moduli space}

Having understood the variation of $P_\xi$ with $\xi$, 
we have to consider the effect of the quotient by $H$ which produces the 
moduli space ${\cal M}_\xi$. We claim that this quotient 
{\em does not modify the wall structure any further}, 
i.e. that the walls in the space of moment map levels  
are the same for ${\cal M}_\xi$ and  $P_\xi$. 

This can be seen as follows. 
If $\xi$ and $\xi'$ both belong to a given chamber for the quotient 
(\ref{first}), then there is a canonical diffeomorphism $\Psi$ between 
$P_\xi$ and $P_{\xi'}$, which can be obtained by considering the 
orbits of the complexification 
$T^\C\approx (\C^*)^r$ of $T$ (the precise statement
\footnote{This is one of the main results of \cite{Kirwan} in the case 
$\xi=0$. The generalization to $\xi \neq 0$ is straightforward and is 
discussed, for example, in \cite{quivers_moduli}.} is as follows. 
Given an orbit $O$ of $T$ inside of ${\cal Z}\cap M_T^{-1}(\xi)$, there 
exists a unique closed orbit ${\cal O}$ 
of $T^\C$ which passes through $O$, and this intersects 
${\cal Z}\cap M_T^{-1}(\xi')$ along an orbit $O'$ 
of $T$ (see Figure 4). Then the isomorphism in question is obtained by sending 
$O$ to $O'$.)

\begin{center}
\input{orbits.pstex_t}

\vskip 0.3 in

\parbox{340pt}{
Figure 4. {\footnotesize The isomorphism between ${\cal M}_\xi$ and 
${\cal M}_{\xi'}$. The right side of the figure depicts two $G$-orbits 
(represented as two `horizontal' sheets) inside of two level 
sets ${\cal Z} \cap M^{-1}(\xi)$ and ${\cal Z} \cap M^{-1}(\xi')$ (not shown) 
for the action of $G$ on the variety of commuting matrices. 
Two $T$-orbits ${\cal O}$, ${\cal O}'$ 
(represented as horizontal lines) inside of these 
$G$-orbits are identified by the map $\Psi$ if they belong to the same orbit 
(the ruled vertical sheet) of the complexified torus  
$T^\C$. The quotient by the action of $T$ (represented by the 
dotted arrow on the right) collapses all $T$-orbits to points, 
thus taking the $G$-orbits into two $H$-orbits inside of the 
varieties 
$P_\xi$, $P_{\xi'}$. These are identified by the map $\Psi$. 
Further quotienting by $H$ (the dotted arrow on the left) gives an isomorphism 
(induced from $\Psi$) between ${\cal M}_\xi$ and ${\cal M}_{\xi '}$.}}
\label{Figure 4.}
\end{center}

Since $T$ is central subgroup of $G$, it follows that $\Psi$ 
commutes with the actions of $H$ on $P_\xi$ and $P_{\xi'}$. 
This assures us that $\Psi$ induces a diffeomorphism between 
${\cal M}_\xi$ and ${\cal M}_{\xi'}$ as we take the quotient by $H$. 
It follows that (the differential type of) ${\cal M}_\xi$ does not 
change as $\xi$ varies in a fixed chamber associated to the action of $T$ 
on ${\cal Z}$. On the other 
hand, if $P_\xi$ does become singular as $\xi$ crosses a wall, 
then ${\cal M}_\xi$ will be singular as well. Therefore, the chamber structure 
for the action of $G$ on  ${\cal Z}$ is the same as that associated to 
the action of $T$ on ${\cal Z}$. 

In conclusion, the wall structure of the space of Fayet-Iliopoulos parameters 
can be determined by examining the first stage of the quotient, which is 
essentially a problem in toric geometry. In the next section we apply this 
method to two low-rank examples, which allows us to locate the flop 
transitions in the 
moduli space of D-branes placed at nonabelian quotient singularities of 
Calabi-Yau threefolds.

\subsection{The toric subgroup of the effective gauge group and its action}

In this subsection, we discuss the first step of the double quotient 
in detail, since, as we argued above, it contains the information we 
wish to extract. 

\subsubsection{The action of $T$ on the quiver data}

Consider a level $\xi$ of the moment map $M$ and the central subgroup
$T= S[\Pi_{\mu=0..r}{U(1)}]/K~~$ of $G$ consisting of the elements of the 
form $(\gamma_0 1_{n_0}..\gamma_r 1_{n_r})\in \Pi_{\mu=0..r}U(n_\mu)$ 
(considered up to the identifications given by $K$) which satisfy the 
constraint $\Pi_{\mu=0..r}{\gamma_\mu^{n_\mu}}=1$. 
Since $n_0=1$, we can solve this constraint for $\gamma_0$ in order to 
present $T$ as:
\be
\label{presentT}
T\approx U(1)^r/C~~,
\ee
i.e. the set of $r$-tuples $(\gamma_1..\gamma_r)\in U(1)^r$ modulo the central 
subgroup $C\approx \Z_n$ given by:
\be
C=\{(\eta...\eta)\in U(1)^r|\eta^n=1\}~~.
\ee
This group acts on the quiver data $\phi^{(\nu\mu)}_\alpha$ through the 
action induced from that of $G$:
\be
\phi^{(\nu\mu)}_{\alpha}\rightarrow e^{i(s_\nu-s_\mu)}
\phi^{(\nu\mu)}_{\alpha}~~,
\ee
where we wrote $\gamma_\mu=e^{is_\mu}$ for all $\mu=0..r$. Using 
$s_0=-\sum_{\mu=1..r}{n_\mu s_\mu}~~(\mbox{mod~} 2\pi)$ we obtain the 
action of $U(1)^r/C$:
\be
\label{toric_action}
\phi^{(\nu\mu)}_{\alpha}\rightarrow e^{il_{\nu\mu}^{(\lambda)} s_\lambda}
\phi^{(\nu\mu)}_{\alpha}~~(\lambda =1..r)~,
\ee
where $l_{\nu\mu}^{(\lambda)}=(1-\delta_\nu^0)(1-\delta_\mu^0)
(\delta_\nu^\lambda -\delta_\mu^\lambda )-
\delta_\nu^0(1-\delta_\mu^0)[n_\lambda+\delta_\mu^\lambda]  
+\delta_\mu^0(1-\delta_\nu^0)
[n_\lambda+\delta_\nu^\lambda]$. 

\subsubsection{The moment map for the toric action}

If ${\cal Q}$ denotes the 
vector space of all quiver data $\{\phi^{(\nu\mu)}_\alpha\}$, then the 
variety ${\cal Z}$ of commuting matrices is realized as a subvariety of 
${\cal Q}$ 
given by the polynomial equations which characterize the critical points of 
the superpotential $W$. Since the superpotential is gauge-invariant, 
${\cal Z}$ is stable under the action of $G$ and this gives a K\"{a}hlerian 
action of $G$ on ${\cal Z}$ induced by the action on ${\cal Q}$. The moment
map for this action is simply the restriction of $M$ to ${\cal Z}$. 

The Lie algebra ${\bf t}=u(1)^{\oplus r}$ of  $T=U(1)^r/C$ is
embedded as a central subalgebra of the Lie algebra 
${\bf g}$ of $G$ via the map:
\be
j(s_1...s_r)=(s_0 1_{n_0},s_1 1_{n_1},...,s_r 1_{n_r})~~,
\ee
with $s_0:=-\sum_{\mu=1..r}{n_\mu s_\mu}$.
The moment map $M_T:{\cal Q}\rightarrow {\bf t}$ 
for the action of $T$ on ${\cal Q}$ is related to $M$ via:
\be
\label{rel}
Tr(M_T(\Phi)s)=Tr(M(\Phi)j(s)), \mbox{~for all~} 
s=(s_1..s_r)\in u(1)^{\oplus r}=\R^r~~.
\ee
Writing $M(\Phi)=\oplus_{\mu=0..r}{M_\mu(\Phi)}$ as in equation 
(\ref{Mmu}) and $M_T(\Phi)=\oplus_{\lambda=1..r}{(M_T)_\lambda(\Phi)}$, with 
$(M_T)_\lambda(\Phi)\in u(1)$, condition (\ref{rel}) gives:
\be
\label{MT}
(M_T)_\lambda(\Phi)=Tr(M_\lambda(\Phi))-n_\lambda Tr(M_0(\Phi))~~.
\ee
In this presentation, the central levels $\xi'_\lambda$ of $M_T$ are related 
to $\xi_\mu$ by:
\be
\label{xi'}
\xi'_\lambda=n_\lambda(\xi_\lambda-\xi_0)~~.
\ee
We will use the presentation (\ref{xi'}) when discussing the examples of 
Section 4. Clearly this trivial reparametrization does not affect any of the 
considerations of the previous subsection: the space $\R^r(\xi')$ has a phase 
stucture obtained from that of $\R^r(\xi)$ by the linear transformation 
(\ref{xi'}).

One can simplify the expression (\ref{MT}) by noticing that:
\be
tr M_\mu=\sum_{\nu,~a_{\nu\mu}\neq 0}{\sum_{\alpha=1..a_{\nu\mu}}
{||\phi^{(\mu\nu)}_\alpha ||^2}}
-\sum_{\nu,~a_{\mu\nu}\neq 0}
{\sum_{\alpha=1..a_{\mu\nu}}
{||\phi^{(\nu\mu)}_\alpha ||^2}}~~,
\ee
where $||\phi^{(\nu\mu)}_\alpha||^2=
\sum_{\scriptsize \begin{array}{c}i=1..n_\mu\\j=1..n_\nu\end{array}}
{|(\phi^{(\nu\mu)}_\alpha)_{ji}|^2}$ (in some orthonormal bases of 
$V_\mu,~V_\nu$) is the usual operator norm of 
$\phi^{(\nu\mu)}_\alpha$. Then a simple computation shows that (\ref{MT}) 
is equivalent with:
\be
\label{toricmoment}
(M_T)_\lambda=\sum_{\scriptsize 
\begin{array}{c}\nu,\mu=0..r\\(a_{\mu\nu}\neq 0)
\end{array}}{l_{\nu\mu}^{(\lambda)} 
\sum_{\alpha=1..a_{\mu\nu}}{||\phi^{(\nu\mu)}_\alpha||^2}}~~(\lambda=1..r),
\ee
which is the standard form \cite{Audin} of the moment map  
for the action (\ref{toric_action}) of the torus $U(1)^r$ on the vector space 
${\cal Q}$.

\section{Examples}

\subsection{Finite subgroups of $SU(3)$}

The finite subgroups of $SU(3)$ fall into 3 series \cite{Fulton}:

(a)A finite series consisting of 6 `crystal-like' groups 
$\Sigma_1...\Sigma_6$, of which the minimal order is $|\Sigma_1|=60$.

(b)Two infinite series of `dihedral-like' groups, denoted by:

(a)$\Delta_1(3n^2)$ (of order $3n^2$), with $n$ any positive integer

(b)$\Delta_2(6n^2)$ (of order $6n^2$), with $n$ any positive {\em even} 
integer. 

For computational reasons, we are interested in subgroups of low rank. 
The lowest ranks are attained by  
$\Delta_1(3)~(n=1)$, $\Delta_1(12)~(n=2)$ and $\Delta_2(24)~(n=2)$.
However, $\Delta_1(3)$ is isomorphic to $A_3$ 
(the alternating group on 3 letters), which is abelian, so we will only 
consider the subgroups $\Delta_1(12)$ and $\Delta_2(24)$. It should be noted 
that all of our statements below apply at the level of classical field theory 
only. In this paper, we do not consider quantum-mechanical modifications of 
the moduli space\footnote{In particular, we neglect quantum anomalies, which 
were recently shown \cite{MP} to be sometimes present in the field theory 
of D-branes placed at orbifold singularities.}

\subsection{The case $\Gamma=\Delta_1(12)$}

The subgroup $\Delta_1(12)$ is given by the following 12 elements
\footnote{It is not hard to see that this group is isomorphic with the 
point group of a regular tetrahedron, i.e. the subgroup of $SO(3)$ which 
leaves such a tetrahedron invariant.}:
\bea
\label{matrices}
A(p,q)&=&\left[\begin{array}{ccc}(-1)^p&0&0\\0&(-1)^q&0\\0&0&(-1)^{p+q}\end{array}\right] \nn~~\\
C(p,q)&=&\left[\begin{array}{ccc}0&0&(-1)^p\\(-1)^q&0&0\\0&(-1)^{p+q}&0\end{array}\right] ~~\\
E(p,q)&=&\left[\begin{array}{ccc}0&(-1)^p&0\\0&0&(-1)^q\\(-1)^{p+q}&0&0\end{array}\right] \nn~~.
\eea
Since $\Delta_1(12)$ has 12 elements, the unprojected D-brane theory has 
a $U(12)$ gauge group and $3\times 12\times 12=432$ complex fields. 
The presence of such a large field content even in the case of the lowest 
order nonabelian orbifold group shows that it is difficult to 
analyze the projection conditions and the variety of commuting matrices 
without making use of the systematic approach we developed in section 2. 
Following that approach, we first note that our orbifold  group has 4 
irreducible representations, which we denote by $R_0$,$R_1$,$R_2$ and $R_3$. 
The representations $R_0,R_1,R_2$ are 1-dimensional (with $R_0$ the 
trivial representation) while $R_3$ is the 3-dimensional defining 
representation induced
by the embedding of $\Delta_1(12)$ in $SU(3)$. We will chose the action of 
$\Delta$ on $\C^3$ to be given by $Q=R_3$. 

The characters of the 4 irreducible representations are given by:
$$
\hbox{
\vbox{\offinterlineskip \tabskip=0pt
\halign{
#&
\vrule height 10pt depth 5pt
\enskip\hfil$#$\hfil\enskip\vrule &
\enskip\hfil$#$\hfil\enskip\vrule &
\enskip\hfil$#$\hfil\enskip\vrule &
\enskip\hfil$#$\hfil\enskip\vrule &
\enskip\hfil$#$\hfil\enskip\vrule \cr\tablerule&
{\rm irrep}&A(p,q)&C(p,q)&E(p,q)\cr \tablerule &
R_0&1&1&1 \cr \tablerule &
R_1&1&\sigma &\sigma^2 \cr \tablerule &
R_2&1&\sigma^2&\sigma \cr \tablerule &
R_3&\phi(p,q)&0&0 \cr \tablerule 
}}}
$$
where $\phi(p,q):=(-1)^p+(-1)^q+(-1)^{p+q}$ and 
$\sigma:=e^{\frac{2\pi i}{3}}$ is the primitive cubic root of unity.

\subsubsection{Branching rules and the quiver}

Use of the characters given above establishes the branching rules:
\bea
Q\otimes R_0&\approx &Q~~\nn\\
Q\otimes R_1&\approx &Q~~\nn\\
Q\otimes R_2&\approx &Q~~\nn\\
Q\otimes R_3&\approx&
Q\otimes Q\approx R_0\oplus R_1 \oplus R_2 \oplus (R_3\otimes R_3)~~,
\eea
which lead to the McKay quiver depicted below. Note that the McKay 
coefficients are symmetric: $a_{\mu \nu}=a_{\nu \mu}$, a reflection 
of the fact that the defining representation $Q$ is self-dual, $Q^*\approx Q$. 
Since each arrow, respectively loop of the quiver corresponds to $3$, 
respectively $9$ complex fields, only $36$ out of the initial $432$ 
fields survive the projection conditions. 

\

\iffigs
$$\vbox{\hskip 1.5 in \hbox{\epsfxsize=5cm\epsfbox{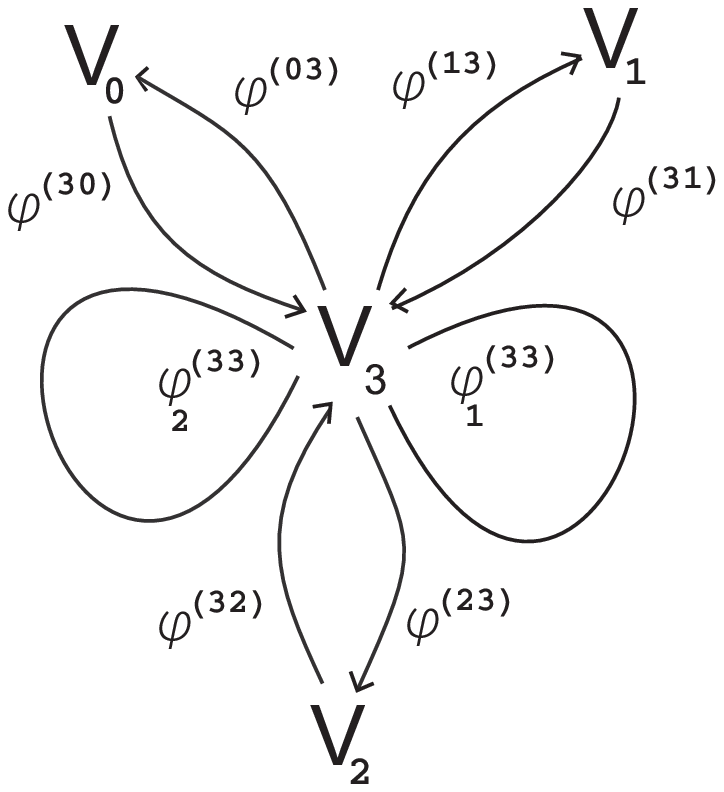}}

\hskip 1.5 in \hbox{Figure 5. {\footnotesize The quiver for  
$\Delta_1(12)$.}}}$$
\fi

\

\subsubsection{Clebsch-Gordan coefficients and the superpotential}

We choose the characters and matrices of the previous subsection as our 
fiducial matrix form of the irreducible representations $R_0..R_3$. 
With these conventions, the Clebsch-Gordan coefficients can be computed 
from the information given above by the method of projectors 
(a brief review of this method is given in Appendix 2). 
For the decompositions 
$Q\otimes R_\mu\approx Q ~(\mu=0..2)$, they are given by:
\be
C^{3i}_{3j,\mu 1}=a^{(\mu)}_i\delta_{ij}~~,
\ee
where $a^{(\mu)}_i (\mu=0..2,i=1..3)$ are the elements of the matrix:
\be
a=\left[\begin{array}{ccc}
1 & 1 & 1\\
1 & \sigma & \sigma^2\\
\sigma^2 & \sigma & 1
\end{array}\right]~~,
\ee
(where $\mu$ is the line index and $i$ is the column index). 
The Clebsch-Gordan coefficients for the decomposition 
$Q\otimes Q\approx R_0\oplus R_1 \oplus R_2 \oplus R_3\otimes R_3$ fall into 
two types: those associated to the $R_\mu ~(\mu=0..2)$ terms, which we denote 
by $C^{\mu 1}_{3 i,3 j}=U^{(\mu)}_{ij}$; and those associated to the 
$R_3\otimes R_3$ term, which we denote by 
$C^{3 k,\alpha}_{3 i,3 j}:=V^{(k \alpha)}_{ij}$. Here $i,j=1..3$ index the 
elements $e^{(3)}_i\otimes e^{(3)}_j \in Q\otimes Q$, with $e^{(\mu)}_i$ the 
fiducial basis of $Q=R_3$, while 
$\alpha, k=1..3$ index a fiducial basis $e^{(3,\alpha)}_k$ of 
$R_3\otimes R_3$, 
with $\alpha$ the multiplicity index. The $R_\mu$ terms ($\mu=0..2)$ 
have fiducial bases given by one element $e^{(\mu)}_1$, since they are 
one-dimensional. With these notations, 
the Clebsch-Gordan coefficients for the decomposition of $Q\otimes Q$ are:
\bea
U^{(1)}=\frac{1}{\sqrt{3}} I_3 ~~~~~~~~~~~&
U^{(2)}=\frac{1}{\sqrt{3}}\left[\begin{array}{ccc}\sigma &0&0\\0&1&0\\0&0&\sigma^2
\end{array}\right] &
U^{(3)}=\frac{1}{\sqrt{3}}\left[\begin{array}{ccc}1&0&0\\0&\sigma&0\\0&0&\sigma^2
\end{array}\right] \nn\\
V^{(1,1)}=\left[\begin{array}{ccc}0&0&0\\0&0&0\\0&1&0\end{array}\right] &
~~~V^{(2,1)}=\left[\begin{array}{ccc}0 &0 &1\\0&0&0\\0&0&0\end{array}\right] &
~~~V^{(3,1)}=\left[\begin{array}{ccc}0 &0 &0\\1&0&0\\0&0&0\end{array}\right] \\
V^{(1,2)}=\left[\begin{array}{ccc}0 &0 &0\\0&0&1\\0&0&0\end{array}\right] &
~~~V^{(2,2)}=\left[\begin{array}{ccc}0 &0 &0\\0&0&0\\1&0&0\end{array}\right] &
~~~V^{(3,2)}=\left[\begin{array}{ccc}0 &1 &0\\0&0&0\\0&0&0\end{array}\right]~\nn~,
\eea
where we used matrix notation: $U^{(\mu)}:=(U^{(\mu)}_{ij})_{i,j=1..3}$, 
$V^{(k,\alpha)}:=(V^{(k,\alpha)}_{ij})_{i,j=1..3}$. 

Applying the quiver Feynman rules gives the superpotential:
\be
W=-3 K^{(\mu)}_\alpha Tr[\phi^{(33)}_{\alpha}\phi^{(3\mu)}\phi^{(\mu 3)}]-
L_{\alpha \beta \gamma}
Tr[\phi^{(33)}_\gamma\phi^{(33)_\beta}\phi^{(33)}_{\alpha}]~~,
\ee
where the cubic couplings $K^{(\mu)}_\alpha$ and $L_{\alpha \beta \gamma}$
($\mu=0..2, \alpha,\beta,\gamma=1,2$) are given by:
\bea
K^{(\mu)}_\alpha=\epsilon_{mnl}a^{(\mu)}_m U^{(\mu)}_{nk} V^{(k\alpha)}_{lm}~~~\\
L_{\alpha \beta \gamma}=
\epsilon_{mnl}V^{(i\alpha)}_{mj}V^{(j\beta)}_{nk}V^{(k\gamma)}_{li}~~.
\eea
Direct computation shows that $K^{(\mu)}_\alpha$ are the components 
of the 3 by 2 matrix:
\be
K=\left[\begin{array}{cc}-1&1\\-\sigma^2&1\\-\sigma&1\end{array}\right]~~,
\ee
while the only nonzero components of $L_{\alpha \beta \gamma}$ are:
\bea
L_{111}=3~~~~~\\
L_{222}=-3~~.
\eea

\subsubsection{The variety of commuting matrices}

It is convenient to denote the quiver data by:
\bea
x=\phi^{(03)}, & y=\phi^{(13)}, & z=\phi^{(23)},\nn\\ 
X=\phi^{(30)}, &Y=\phi^{(31)}, & Z=\phi^{(32)},\nn\\
u=\phi^{(33)}_1, & v=\phi^{(33)}_2~~.\nn\\  
\eea
With respect to the $GL(3)$ action, the $x,y,z$ are $3\times 1$ matrices
transforming as covectors, the $X,Y,Z$ are $1\times 3$ matrices transforming 
as vectors, and the $u,v$  are $3\times 3$ matrices transforming as tensors 
of type $(1,1)$.
With this notation, the superpotential becomes:
\be
W=-3[x(v-u)X+y(v-\sigma^2 u)Y+z(v-\sigma u)Z+Tr(u^3)-Tr(v^3)]~~.
\ee
Differentiating $W$ leads to the F-flatness constraints:
\bea
\label{Fexplicit}
(v-u)X=0, & (v-\sigma^2u)Y=0, & (v-\sigma u)Z=0 \\
x(v-u)=0, & y(v-\sigma^2 u)=0, & z(v-\sigma u)=0 \\
3u^2=Xx+\sigma^2Yy +\sigma Zz \\
3v^2=Xx+Yy+Zz~~,
\eea
whose solution set is the variety of commuting matrices ${\cal Z}$.

\subsubsection{The toric part of the quotient and the enlarged K\"{a}hler cone}

The effective gauge group is given by:
\be
\label{group1}
G=U(1)^3\times U(3)/U(1)_{\diag}\approx S[U(1)^3\times U(3)]/K~~,
\ee
where $S[U(1)^3\times U(3)]$ is the group of quadruples 
$(g_0,g_1,g_2,g_3) \in U(1)^3\times U(3)$ such that $g_0g_1g_2detg_3=1$ 
and $K\approx\Z_6$ is its central subgroup:
\be
K=\{(\eta,\eta,\eta,\eta 1_3)|\eta^6=1\}~~.
\ee
$G$ has a central subgroup $T=T_0/K$, where: 
\be
T_0=\{(g_0,g_1,g_2,\gamma)\in U(1)^4 | g_0g_1g_2\gamma^3=1\}~~,
\ee
which corresponds to setting $g_3=\gamma 1_3$. We have an isomorphism 
$T\approx U(1)^3/C$, induced by:
\be
(g_1,g_2,\gamma)\rightarrow (g_1^{-1}g_2^{-1}\gamma^{-3},g_1,g_2,\gamma)~~,
\ee
which corresponds to solving for $g_0$ in the constraint 
$g_0g_1g_2\gamma^3=1$. $C\approx \Z_6$ is a central subgroup of 
$U(1)^3$ given by:
\be
C=\{(\eta,\eta,\eta)|\eta^6=1\}~~.
\ee
Writing $g_1:=e^{is_1},g_2:=e^{is_2},\gamma=e^{is_3}$, the subgroup  
$T\approx U(1)^3/\Z_6$ of $G$ acts on the quiver data as 
$\phi^{(\mu3)}\rightarrow e^{il^{(\lambda)}_\mu s_\lambda}\phi^{(\mu3)}$,
$\phi^{(3\mu)}\rightarrow e^{-il^{(\lambda)}_\mu s_\lambda}\phi^{(3\mu)}$,
while leaving $\phi^{(33)}_1$ and $\phi^{(33)}_2$ unchanged, with the charge 
matrix $L:=(l^{(\lambda)}_\mu)_{\lambda=1..3,\mu=0..2}$ given by:
\be
L=\left[\begin{array}{ccc}-1 & 1& 0\\-1&0&1\\-4&-1&-1\end{array}\right]~~
\ee
(where $\lambda$ is the line index and $\mu$ is the column index). Therefore, 
the toric part of the double quotient has the form 
$(\C^{18}-Z_{\xi'})/(\C^*)^3$.
The charge matrix $L$ is invertible with inverse:
\be
\label{inverse}
L^{-1}
=\frac{1}{6}
\left[\begin{array}{ccc} -1&-1&-1\\\noalign{\medskip}5&-1&
-1\\\noalign{\medskip}-1&5&-1\end{array}\right ]\in 
GL(3,\Q)~~.
\ee

The moment map for this (effective)
$U(1)^3/\Z_6$ action is given by:
\be
(M_T)_\lambda=
\sum_{\mu=0..2}{l_\mu^{(\lambda)}[||\phi^{(\mu 3)}||^2-||\phi^{(3\mu)}||^2]}.
\ee
Picking a level $\xi'=(\xi'_\lambda)_{\lambda=1..3}\in \R^3$, 
we can solve the moment map 
equations $(M_T)_\lambda=\xi'_\lambda$ as:
\be
\label{level_set}
||\phi^{(\mu 3)}||^2-||\phi^{(3\mu)}||^2=t_\mu~~ (\mu=0..2)~~,
\ee
where $t_\mu$ are the components of the real vector $t=L^{-1}\xi'$. 
It is well-known that $\xi'$ is a regular level of the moment map 
if and only if 
the stabilizer of any point in $M^{-1}(\xi')$ is finite.  
The points $x\in M^{-1}(\xi')$ which do not satisfy this condition 
are stabilized by a $U(1)$ subgroup of $G$ and correspond
(via the Higgs mechanism) to classical vacua 
admitting some unbroken gauge symmetry. Therefore, the nonregular values of 
$\xi'$ are precisely those values for which the classical moduli space contains
points of enhanced gauge symmetry. In our case, 
the above criterion immediately shows\footnote{
Introduce the notation $\phi^{(\mu 3)}:=\phi^{(\mu+)},
\phi^{(3 \mu)}:=\phi^{(\mu-)}$ and assume first 
that $t_1,t_2,t_3$ are all nonzero.
Then let $\epsilon_\mu:=sign(t_\mu)\in \{-1,1\}$. If 
$\phi^{(\mu+)},\phi^{(\mu-)}$ satisfy (\ref{level_set}), then we necessarily 
have $||\phi^{(\mu\epsilon_\mu)}||\neq 0$ for all $\mu=0..2$. 
Such a point is fixed by an element $(\sigma_1..\sigma_3)\in U(1)^3$ 
if and only if 
$\Pi_{\lambda=1..3}\sigma_{\lambda}^{\epsilon_\mu l^{(\lambda)}_\mu}=1$ 
for all 
$\mu=0..2$. Let $L^{-1}:=\frac{1}{6}A$, where $A$ is the integral matrix 
in equation (\ref{inverse}). Then $LA=6$, and taking products of the above 
equations for $\sigma$ shows that $\sigma_\lambda^6=1$ for all $\lambda=1..3$.
Hence the stabilizer of any point in the level set considered is a subset of 
$(\C_6)^3$(with $\C_6$ the group of roots of unity of order 6), and therefore 
is finite. On the other hand, if some $t_\mu$ is zero, then one can easily 
construct a continuous subgroup of $T$ which fixes some point of the 
associated level set.}
that $\xi'$ is a regular level of $M$ if and only if 
$\Pi_{\mu=0..2}{t_\mu}\neq 0$. 
The set $\Pi_{\mu=0..2}{t_\mu}=0$ of singular 
levels coincides with the union of the coordinate planes in the space 
$\R^3(t)$ of all values of $t$. This divides $\R^3(t)$ into the octants
$\Sigma_\epsilon$ ($\epsilon:=(\epsilon_1,\epsilon_2,\epsilon_3)$, with 
$\epsilon_i=-1$ or $1$), which are just the cones 
$\Sigma_\epsilon=\langle\epsilon_1e_1,\epsilon_2e_2,
\epsilon_3e_3\rangle_{\R^+}$ 
spanned by the vectors $(\epsilon_ie_i)_{i=1..3}$, with $(e_i)_{i=1..3}$ the 
canonical basis of $\R^3$. 
Since $\xi'=Lt$, the space $\R^3(\xi')$ of values of $\xi'$ is similarly divided 
into the 8 cones 
$\sigma_\epsilon=\langle\epsilon_1u_1,\epsilon_2u_2,
\epsilon_3u_3\rangle_{\R^+}$ ($\epsilon_i=\pm 1$), where 
$u_i=Le_i$ are the column vectors of $L$:
\bea
u_1:=\left[\begin{array}{c}-1\\-1\\4\end{array}\right]&
u_2:=\left[\begin{array}{c}1\\0\\-1\end{array}\right]&
u_3:=\left[\begin{array}{c}0\\1\\-1\end{array}\right]~~.
\eea
These cones are the GIT chambers of the ambient space in our situation. 
The critical values of the moment map are the points of the walls:
\bea
W_1=\langle u_1,u_2\rangle_{\R}~,& W_2=\langle u_2,u_3\rangle_{\R}~,
& W_3=
\langle u_3,u_1\rangle_{\R}~~,
\eea
which give a system of 3 planes in $\R^3(\xi')$, intersecting at the origin 
and along the vectors $u_1,u_2,u_3$ (see Figure 6). 

\

\iffigs
$$\vbox{\hskip 1.3 in \hbox{\epsfxsize=7cm\epsfbox{cones12.ps}}

\hskip 1.0 in \hbox{Figure 6. 
{\footnotesize The enlarged K\"{a}hler cone for $\C^3/\Delta_1(12)$.}}}$$
\fi

\

If $\xi'$ belongs to the cone $\sigma_\epsilon$~
($\epsilon=(\epsilon_0,\epsilon_1,\epsilon_2)\in \{-1,1\}\times
\{-1,1\}\times \{-1,1\}$) then the exceptional set 
$Z_\xi'$ is given by:
\be
Z_\epsilon=\cup_{\mu=0..2}{Z^{(\mu)}_{\epsilon_\mu}}=\{\phi \in {\cal Q}|
\phi^{(0,\epsilon_0)}=\phi^{(1,\epsilon_1)}=\phi^{(2,\epsilon_2)}=0\}~~,
\ee
where:
\be
Z^{(\mu)}_{\epsilon_\mu}=\{\phi \in {\cal Q}|\phi^{(\mu,\epsilon_\mu)}=0\}~,
\ee
(we use the notation $\phi^{(\mu 3)}:=\phi^{(\mu+)},
\phi^{(3 \mu)}:=\phi^{(\mu-)}$).
It is not hard to see that the exceptional sets $Z_\epsilon$ intersect 
the variety of commuting matrices ${\cal Z}$ along different loci, 
so that none of the walls is `projected out' (or identified with another wall) 
upon restricting to ${\cal Z}$. The defining equations (\ref{Fexplicit}) of 
the variety ${\cal Z}$ can be solved (on a dense open subset) 
\footnote{This can be achieved, for example,by 
using linear algebra arguments related to completeness relations for 
the left and right eigenvectors of the 
$3$ by $3$ matrices $v-u,~v-\sigma^2u$ and $v-\sigma u=0$.},
which allows one to show that the quotient ${\cal Z}/G_\C$ is complex 
3-dimensional, as expected.

\subsection{The case $\Gamma:=\Delta_2(24)$}

The subgroup $\Delta_2(24)$ is given by the  12 matrices 
$A(p,q),C(p,q),E(p,q)$ of (\ref{matrices}) together with the following 12 elements:
\bea
B(p,q)&=&\left[\begin{array}{ccc}(-1)^p&0&0\\0&0&(-1)^q\\0&(-1)^{p+q}&0\end{array}\right]\nn~~\\
D(p,q)&=&\left[\begin{array}{ccc}0&(-1)^p&0\\(-1)^q&0&0\\0&0&(-1)^{p+q}\end{array}\right]~~\\
F(p,q)&=&\left[\begin{array}{ccc}0&0&(-1)^p\\0&(-1)^q&0\\(-1)^{p+q}&0&0\end{array}\right]\nn~~,\\
\eea
where $p,q=0,1$. 

	This group is isomorphic to the more familiar symmetric group on 
four letters.  The unprojected D-brane theory has 
a $U(24)$ gauge group and $3\times 24\times 24=1,728$ complex fields.  Proceeding as before, we note that our orbifold group has five irreducible representations, 
which we denote by $R_0$, $R_1$, $R_2$, $R_3$ and
$R_4$.  The representations $R_0$, $R_1$, and $R_2$ are induced from
representations of the symmetric group on three letters via the isomorphism 
$\Delta_2(24)/\{A(p,q)\}\approx S_3$.~~$R_0$ is the 
trivial representation, $R_1$ is the one dimensional sign representation, 
and $R_2$ is the two dimensional triangle representation.  
$R_3$ is the three dimensional
defining representation induced by the embedding of $\Delta_2(24)$ in
$SU(3)$ via the matrices given above.  $R_4$ is the three dimensional 
representation whose matrices are identical to those of $R_3$ except that 
$B(p,q)$, $D(p,q)$, and $F(p,q)$ are multiplied by $-1$.  $R_4$ is not special
unitary, so we only consider $Q=R_3$ as the action of $\Delta_2(24)$ on
$\C^3$.

	$\Delta_2(24)$ has five conjugacy classes:
\bea
1C1 &=& A(0,0)~~\nn\\ 3C2 &=& A(0,1),A(1,0),A(1,1)~~\nn\\ 8C3 &=&
C(p,q),E(p,q),~p,q\in0,1~~\nn\\ 6C4 &=&
B(0,0),B(0,1),D(0,1),D(1,0),F(0,0),F(1,0)~~\nn\\ 6C5 &=&
B(1,0),B(1,1),D(0,0),D(1,1),F(0,1),F(1,1)~~,\nn\\
\eea
where we use the 
standard group-theoretic notation $iCj$ for conjugacy classes, with $i$ the 
number of elements of the class and $j$ its label. 
The character table is:
$$
\hbox{
\vbox{\offinterlineskip \tabskip=0pt
\halign{
#&
\vrule height 10pt depth 5pt
\enskip\hfil$#$\hfil\enskip\vrule &
\enskip\hfil$#$\hfil\enskip\vrule &
\enskip\hfil$#$\hfil\enskip\vrule &
\enskip\hfil$#$\hfil\enskip\vrule &
\enskip\hfil$#$\hfil\enskip\vrule &
\enskip\hfil$#$\hfil\enskip\vrule \cr\tablerule&
{\rm irrep}&1C1&3C2&8C3&6C4&6C5\cr \tablerule & R_0&1&1&1&1&1 \cr
\tablerule & R_1&1&1&1&-1&-1 \cr \tablerule & R_2&2&2&-1&0&0 \cr
\tablerule & R_3&3&-1&0&1&-1 \cr \tablerule & R_4&3&-1&0&-1&1 \cr
\tablerule }}}
$$

\subsubsection{Branching rules and the quiver}

Use of the characters above establishes the following branching rules:
\bea
\\
Q\otimes R_0 &\approx & Q~~\nn\\ 
Q\otimes R_1 &\approx & R_4~~\nn\\ 
Q\otimes R_2 &\approx & Q\oplus R_4~~\\ 
Q\otimes R_3 &\approx & R_0\oplus R_2\oplus Q\oplus R_4~~\nn\\ 
Q\otimes R_4 &\approx & R_1\oplus R_2 \oplus Q\oplus
R_4~~,\nn
\eea
which lead to the McKay quiver depicted below. The McKay
coefficients are symmetric: $a_{\mu \nu}=a_{\nu \mu}$, a property due to 
self-duality of the defining representation $Q$. The quiver is depicted in 
the figure below. There are $72$ complex 
fields surviving the projection conditions: $3+3=6$ from the maps 
$\phi^{(30)}$ and $\phi^{(03)}$, $3+3=6$ from the maps 
$\phi^{(41)}$ and $\phi^{(14)}$, $9$ from $\phi^{(33)}$, 
$9$ from $\phi^{(44)}$, $9+9=18$ from $\phi^{(43)}$ and $\phi^{(34)}$, 
$6+6=12$ from $\phi^{(23)}$ and $\phi^{(32)}$ and $6+6=12$ from $\phi^{(24)}$ 
and $\phi^{(42)}$.

\

\iffigs
$$\vbox{\hskip 1.0 in \hbox{\epsfxsize=10cm\epsfbox{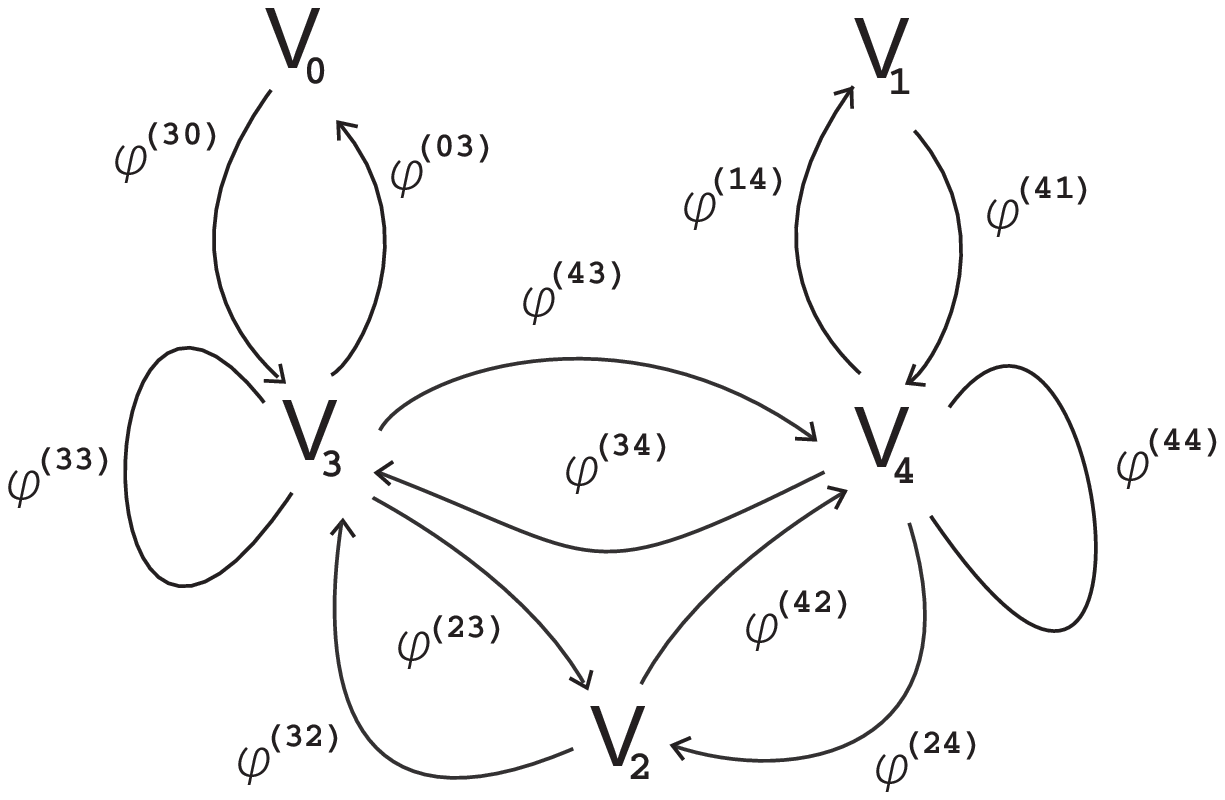}}

\hskip 1.5 in \hbox{\footnotesize Figure 7. The quiver for $\Delta_2(24)$.}}$$
\fi

\

\subsubsection{Clebsch-Gordan coefficients and the superpotential}

The Clebsch-Gordan coefficients can be computed from the information
given above by the method of projectors.  In the case of $\Delta_2(24)$
there are no multiplicities $\alpha$.  Our decompositions reduce
to the simpler form:
\be
\label{dec}
Q\otimes R_\nu=\oplus_\mu{ R_\mu}~~,
\ee
where $\mu$ runs over a subrange of $0..r$. Again we chose the characters 
and matrices given above as the fiducial forms of our ireducible 
representations. 
Let ${n_\mu}$ be the dimension of ${R_\mu}$.  Then for each ${R_\mu}$
appearing in (\ref{dec}), we define ${n_\mu}$ matrices 
${C^{(\mu i)}_\nu}$ (of type $3\times n_\nu$)
by: $(C^{(\mu i)}_\nu)_{jk}:=C^{\mu i}_{Qj,\nu k}$.
These  matrices can be computed by the method of projectors, with the result:
\bea
{C^{(31)}_{0}}=[1,0,0] & {C^{(32)}_{0}}=[0,1,0] & {C^{(33)}_{0}}=[0,0,1]\nn\\
\eea
\bea
{C^{(41)}_{1}}=\left[1,0,0\right] & {C^{(42)}_{1}}=\left[0,1,0\right] & {C^{(43)}_{1}}=\left[0,0,1\right]\nn\\
\eea
\bea
{C^{(31)}_{2}}=\left[\begin{array}{ccc}
1 & 0\\
0 & 0\\
0 & 0
\end{array}\right] &
{C^{(32)}_{2}}=\frac{1}{2}\left[\begin{array}{ccc}
0 & 0\\
-1 & -{\sqrt3} \\
0 & 0
\end{array}\right] &
{C^{(33)}_{2}}=\frac{1}{2}\left[\begin{array}{ccc}
0 & 0\\
0 & 0\\
-1& {\sqrt3}
\end{array}\right]\nn\\
{C^{(41)}_{2}}=\left[\begin{array}{ccc}
0 & 1\\
0 & 0\\
0 & 0
\end{array}\right] &
{C^{(42)}_{2}}=\frac{1}{2}\left[\begin{array}{ccc}
0 & 0\\
{\sqrt3} & -1\\
0 & 0
\end{array}\right] &
{C^{(43)}_{2}}=\frac{1}{2}\left[\begin{array}{ccc}
0 & 0\\
0 & 0\\
-{\sqrt3} & -1 
\end{array}\right]\nn\\
\eea
\bea
{C^{(01)}_{3}}=\frac{1}{\sqrt3}\left[\begin{array}{ccc}
1 & 0 & 0 \\
0 & 1 & 0 \\
0 & 0 & 1 
\end{array}\right]  & 
{C^{(21)}_{3}}=\frac{1}{\sqrt6}\left[\begin{array}{ccc}
-2 & 0 & 0\\
0 & 1 & 0\\
0  & 0 & 1
\end{array}\right] & 
{C^{(22)}_{3}}=\frac{1}{\sqrt2}\left[\begin{array}{ccc}
0 & 0 & 0\\
0 & 1 & 0\\
0 & 0 & -1
\end{array}\right]\nn\\
{C^{(31)}_{3}}=\frac{1}{\sqrt2}\left[\begin{array}{ccc}
0 & 0 & 0\\
0 & 0 & -1\\
0  & 1 & 0
\end{array}\right] & 
{C^{(32)}_{3}}=\frac{1}{\sqrt2}\left[\begin{array}{ccc}
0 & 0 & 1\\
0 & 0 & 0\\
-1 & 0 & 0
\end{array}\right] & 
{C^{(33)}_{3}}=\frac{1}{\sqrt2}\left[\begin{array}{ccc}
0 & -1 & 0\\
1 & 0 & 0\\
0  & 0 & 0
\end{array}\right]\nn\\
{C^{(41)}_{3}}=\frac{1}{\sqrt2}\left[\begin{array}{ccc}
0 & 0 & 0\\
0 & 0 & 1\\
0  & 1 & 0
\end{array}\right] & 
{C^{(42)}_{3}}=\frac{1}{\sqrt2}\left[\begin{array}{ccc}
0 & 0 & 1\\
0 & 0 & 0\\
1  & 0 & 0
\end{array}\right] &
{C^{(43)}_{3}}=\frac{1}{\sqrt2}\left[\begin{array}{ccc}
0 & 1 & 0\\
1 & 0 & 0\\
0  & 0 & 0
\end{array}\right]\nn\\
\eea
\bea
{C^{(11)}_{4}}=\frac{1}{\sqrt3}\left[\begin{array}{ccc}
1 & 0 & 0 \\
0 & 1 & 0 \\
0 & 0 & 1 
\end{array}\right] &
{C^{(21)}_{4}}=\frac{1}{\sqrt2}\left[\begin{array}{ccc}
0 & 0 & 0\\
0 & 1 & 0\\
0  & 0 & -1
\end{array}\right] &
{C^{(22)}_{4}}=\frac{1}{\sqrt6}\left[\begin{array}{ccc}
2 & 0 & 0\\
0 & -1 & 0\\
0 & 0 & -1 
\end{array}\right]\nn\\
{C^{(31)}_{4}}=\frac{1}{\sqrt2}\left[\begin{array}{ccc}
0 & 0 & 0\\
0 & 0 & 1\\
0  & 1 & 0
\end{array}\right] &
{C^{(32)}_{4}}=\frac{1}{\sqrt2}\left[\begin{array}{ccc}
0 & 0 & 1\\
0 & 0 & 0\\
1 & 0 & 0
\end{array}\right] &
{C^{(33)}_{4}}=\frac{1}{\sqrt2}\left[\begin{array}{ccc}
0 & 1 & 0\\
1 & 0 & 0\\
0  & 0 & 0
\end{array}\right]\nn\\
{C^{(41)}_{4}}=\frac{1}{\sqrt2}\left[\begin{array}{ccc}
0 & 0 & 0\\
0 & 0 & -1\\
0  & 1 & 0
\end{array}\right] &
{C^{(42)}_{4}}=\frac{1}{\sqrt2}\left[\begin{array}{ccc}
0 & 0 & 1\\
0 & 0 & 0\\
-1  & 0 & 0
\end{array}\right] &
{C^{(43)}_{4}}=\frac{1}{\sqrt2}\left[\begin{array}{ccc}
0 & -1 & 0\\
1 & 0 & 0\\
0  & 0 & 0
\end{array}\right]\nn\\
\eea
Applying the quiver Feynman rules gives the superpotential:
\bea
W={3\over{\sqrt2}}\left[Tr(\phi^{(33)}\phi^{(33)}\phi^{(33)}) + 
Tr(\phi^{(44)}\phi^{(44)}\phi^{(44)})\right]\nn~~\\
-{3\sqrt6}\left[Tr(\phi^{(30)}\phi^{(33)}\phi^{(03)}) + 
Tr(\phi^{(41)}\phi^{(44)}\phi^{(14)})\right]\nn~~\\
+{3\sqrt3}\left[Tr(\phi^{(42)}\phi^{(44)}\phi^{(24)}) - 
Tr(\phi^{(32)}\phi^{(33)}\phi^{(23)})\right]\nn~~\\
+9\left[Tr(\phi^{(32)}\phi^{(43)}\phi^{(24)}) + 
Tr(\phi^{(42)}\phi^{(34)}\phi^{(23)})\right]\nn~~\\
+{9\over\sqrt2}\left[Tr(\phi^{(43)}\phi^{(44)}\phi^{(34)}) +
Tr(\phi^{(34)}\phi^{(33)}\phi^{(43)})\right]\nn~~.\\
\eea

\subsubsection{The variety of commuting matrices}

Introducing the following notation:
\bea
x=\phi^{(30)},& y=\phi^{(41)}, & s=\phi^{(43)},\nn\\ 
X=\phi^{(03)}, &Y=\phi^{(14)},& S=\phi^{(34)},\nn\\
a=\phi^{(32)}, &b=\phi^{(42)},& U=\phi^{(33)},\nn\\
A=\phi^{(23)}, &B=\phi^{(24)},& V=\phi^{(44)},\nn\\
\eea
the superpotential becomes:
\bea
W={3\over{\sqrt2}}\left[Tr(U^3) + Tr(V^3)\right]-
{3\sqrt6}\left[xUX + yVY\right]\nn~~\\
+{3\sqrt3}\left[Tr(bVB)-Tr(aUA)\right]+9\left[Tr(asB) + Tr(bSA)\right]\nn~~\\
+{9\over\sqrt2}\left[Tr(sVS) +Tr(SUs)\right]\nn~~.\\
\eea
Differentiating $W$ leads to the F-flatness constraints:
\bea
{9\over{\sqrt2}}U^2-3{\sqrt6}Xx-3{\sqrt3}Aa+{9\over{\sqrt2}}Ss=0\nn~~\\
{9\over{\sqrt2}}V^2-3{\sqrt6}Yy-3{\sqrt3}Bb+{9\over{\sqrt2}}sS=0\nn~~\\
9Ba+{9\over{\sqrt2}}S(V+U)=0\nn~~\\
9Ab+{9\over{\sqrt2}}s(V+U)=0\nn~~\\
-3{\sqrt3}{\sqrt2}UX=0 & -3{\sqrt3}{\sqrt2}VY=0\nn~~\\
-3{\sqrt3}{\sqrt2}xU=0 & -3{\sqrt3}{\sqrt2}yV=0\nn~~\\
-3{\sqrt3}UA+9sB=0 & 3{\sqrt3}VB+9SA=0\nn~~\\
-3{\sqrt3}aU+9bS=0 & 3{\sqrt3}bV+9as=0\nn,\\
\eea
whose solution set is the variety of commuting matrices ${\cal Z}$.

\subsubsection{The toric part of the quotient and the enlarged Kahler cone}

The effective gauge group is given by:
\bea
G=U(1)^2\times U(2)\times U(3)^2/U(1)_{\diag}\approx S[U(1)^2\times U(2)\times U(3)^2]/K~~,
\eea
where $S[U(1)^2\times U(2)\times U(3)^2]$ is the group of quintuples 
\bea
{(g_0,g_1,g_2,g_3,g_4)} \in U(1)^2\times U(2)\times U(3)^2\\
\eea
such that
\bea
g_0g_1detg_2detg_3detg_4=1\\
\eea
and $K\approx\Z_{10}$ is its central subgroup:
\be
K=\{(\eta,\eta,\eta 1_2,\eta 1_3,\eta 1_3)|\eta^{10}=1\}~~.
\ee
$G$ has a central subgroup $T=T_0/K$, where: 
\be
T_0=\{(g_0,g_1,\delta,\epsilon,\gamma)\in U(1)^5 | g_0g_1\delta^2\epsilon^3\gamma^3=1\}~~,
\ee
which corresponds to setting $g_2=\delta 1_2$, $g_3=\epsilon 1_3$, $g_3=\gamma 1_3$. We have an isomorphism 
$T\approx U(1)^4/C$, given by:
\be
(g_1,\delta,\epsilon,\gamma)\rightarrow (g_1^{-1}\delta^{-2}\epsilon^{-3}\gamma^{-3},g_1,\delta,\epsilon,\gamma)~~,
\ee
which corresponds to solving for $g_0$ in the constraint 
\be
g_0g_1\delta^2\epsilon^3\gamma^3=1~~.
\ee
$C\approx \Z_{10}$ is a central subgroup of $U(1)^4$ given by:
\be
C=\{(\eta,\eta,\eta,\eta)|\eta^{10}=1\}~~.
\ee
Writing $g_\lambda:=e^{is_\lambda}$, for $\lambda  = 1..4$, the subgroup  
$T\approx U(1)^4/\Z_{10}$ of $G$ acts on the quiver data as 

\bea
\phi^{(\mu\nu)}\rightarrow e^{il^{(\lambda)}_{\mu\nu} s_\lambda}\phi^{(\mu\nu)}\nn\\
\phi^{(\nu\mu)}\rightarrow e^{-il^{(\lambda)}_{\mu\nu} s_\lambda}\phi^{(\nu\mu)}\nn\\
\eea
for $\mu\not=\nu$ while leaving $\phi^{(33)}$ and $\phi^{(44)}$ unchanged.
If we relabel the maps $\phi^{(\nu\mu)}$ for $\mu\not=\nu$ as follows:
\bea
\phi^{(30)}=\phi^{(0)} & \phi^{(41)}=\phi^{(1)} & \phi^{(32)}=\phi^{(2)}\nn\\
\phi^{(42)}=\phi^{(3)} & \phi^{(43)}=\phi^{(4)}\nn\\
\eea
we can re-express the group action as
\bea
\label{action_mark}
\phi^{(\rho)}\rightarrow e^{il^{(\lambda)}_{\rho} s_\lambda}\phi^{(\rho)}\nn\\
\eea
and obtain the charge matrix  
$L:=(l^{(\lambda)}_\rho)_{\lambda=1..4,\rho=0..4}$:
\bea
L=\left[\begin{array}{ccccc}
-1 & 0 & 1 & 1& 0\\
-2 &1& 0 & 0 & 0\\
-4 & 0 & -1 & 0 & 1\\
-3 &-1& 0 & -1& -1
\end{array}\right]\nn~~.\\
\eea
(Here $\lambda$ is the row index and $\rho$ is the column index.)  

Define the vector
\be
	\epsilon^{\rho}=[||\phi^{(\rho)}||^2-||\phi^{(\bar \rho)}||^2],
\ee
where $\phi^{(\bar \rho)}$ represents the arrow going in the opposite 
direction of $\phi^{(\rho)}$:
\be
\phi^{(\rho)}:=\phi^{(\mu\nu)}~~\Rightarrow~~ 
\phi^{(\bar \rho)}:=\phi^{(\nu\mu)}\\
\ee

The moment map for our (effective) $U(1)^4/\Z_{10}$ action is given by:
\be
M_\lambda=\sum_{\rho=0..4}{l_\rho^{(\lambda)}\epsilon^{\rho}}.
\ee

	The charge matrix L does not have a left inverse, so we cannot immediately apply the methods of the earlier example.  Instead first pick a level 
$\xi'=(\xi'_\lambda)_{\lambda=1..4}\in \R^4$ and bring the terms containing $\epsilon^4$ in 
each sum to the left hand side of the equation.  Then the moment map equations become:
\be
{\xi'_\lambda}-{l_4^{(\lambda)}\epsilon^4}=
\sum_{\rho=0..3}{l_\rho^{(\lambda)}\epsilon^{\rho}}.
\ee
	The square part of $L$, for ${\rho}=0..3$, denoted ${\bar L}$, does possess an inverse:

\bea
{\bar L}=\left[\begin{array}{cccc}
-1 & 0 & 1 & 1\\
-2 &1& 0 & 0\\
-4 & 0 & -1 & 0\\
-3 &-1& 0 & -1
\end{array}\right]\nn~~\\
\eea

\bea
{{\bar L}^{-1}}={-1\over 10}\left[\begin{array}{cccc}
1 & 1 & 1 & 1\\
2 & -8 & 2 & 2\\
-4 & -4 & 6 & -4\\
-5 & 5 & -5 & 5
\end{array}\right]\nn~~.\\
\eea

	We then obtain the following equations:

\bea
\label{ine}
{-1\over{10}}(\xi'_1+\xi'_2+\xi'_3+\xi'_4)=\epsilon^0\nn\\
{-1\over{10}}(2\xi'_1-8\xi'_2+2\xi'_3+2\xi'_4)=\epsilon^1\nn\\
{-1\over{10}}(-4\xi'_1-4\xi'_2+6\xi'_3-4\xi'_4)=\epsilon^2-\epsilon^4\\
{-1\over{10}}(-5\xi'_1+5\xi'_2-5\xi'_3+5\xi'_4)=\epsilon^3+\epsilon^4.\nn
\eea

	In particular, there are no constraints on $\epsilon^4$.  
We will therefore have a conical singularity at the point in the moduli space 
corresponding to $\phi^{(4)}=\phi^{({\overline 4})}=0$, which is a 
fixed point for a single 
$U(1) \subset U(1)^4$ 
, independent of the 
value of $\xi'$
\footnote{This $U(1)$ subgroup is given by setting 
$s_1=s_3=-s_2=-s_4$ in equation (\ref{action_mark}).}.

A chamber $\Sigma \subset \R^4(\xi')$ is such that for all $\xi' \in \Sigma$, 
a maximal number of $\epsilon^{i} \neq 0$.  
In our case, we can only enforce such a restriction on $\epsilon^0$ and 
$\epsilon^1$ via the first two equations of (\ref{ine}).  
Our chambers will correspond to four cases:

\bea
\begin{array}{ccc}
(I) & \epsilon^0 > 0 & \epsilon^1 > 0\\
(II)& \epsilon^0 > 0 & \epsilon^1 < 0\\
(III) & \epsilon^0 < 0 & \epsilon^1 > 0\\
(IV) & \epsilon^0 < 0 & \epsilon^1 < 0
\end{array}\nn~~.
\eea

The chamber walls are determined by the union of the two hyperplanes obtained 
by setting $\epsilon^0$ and $\epsilon^1$ to zero in the first two equations 
of (\ref{ine}):

\bea
\xi'_0+\xi'_1+\xi'_2+\xi'_3=0\\
2\xi'_0-8\xi'_1+2\xi'_2+2\xi'_3=0~~.
\eea
\noindent The normal vectors of these hyperplanes can be written as 
$a_0 = (-1,-1,-1,-1)$ and $a_1 = (-1,4,-1,-1)$.  The hyperplanes 
intersect along a two-plane spanned by the vectors $(1,0,0,-1)$ and 
$(0,0,1,-1)$.  The four regions of $\R^4(\xi')$ bounded by these hyperplanes 
correspond to the GIT chambers in this example.

	We can  read  the  chamber conditions  in  terms  of the  maps
	$\phi^{\rho}$.  In
	   chamber (I),  both   $\epsilon^0$    and
	$\epsilon^1$ are    strictly  positive.   This    implies that
	$||\phi^{(0)}||^2$  and     $||\phi^{(1)}||^2$   are  nonzero.
	Other choices for the signs  of   $\epsilon^0$ and
	$\epsilon^1$  result  in the remaining   chambers.  Since 
	$\epsilon^4$ cannot be fixed  in such  a manner, we  are
	free to  have    either  or both of   $||\phi^{(4)}||^2$   and
	$||\phi^{(\bar 4)}||^2$ equal  to  zero.  At any point in   the
	moduli  space corresponding both of these $3\times 3$ matrices being 
	zero,
	$18$ of our $72$ complex field variables vanish and the space develops 
	a conical singularity.

\section{Conclusions and directions for further research}

The moduli space of D-branes placed at quotient singularities is an important 
and fascinating subject, with numerous physical and mathematical 
ramifications. 
In this paper, we have focused on only a few aspects of this topic;
clearly much remains to be resolved. 

One of the most obvious problems to address is that of explicitly computing 
the moduli spaces of our theories. This is related to similar problems in the
context of arbitrary supersymmetric field theories in a given dimension 
\cite{SUSYGIT} and to recent efforts \cite{Csaki} 
to systematize the computation 
of classical moduli spaces of supersymmetric field theories with 4 
supercharges. This is a prerequisite, for example, for understanding 
the phenomena of \cite{Seiberg} in a more general context. In fact, the 
phenomenon of topology change is ubiquitous for the moduli space 
of supersymmetric field theories with unbroken $U(1)$ factors of the gauge 
group, and deserves a much better understanding. The determination 
of the classical moduli space can be handled by methods of algebraic 
geometry, and can be reduced to the problem of computing 
{\em invariants} in the sense of constructive invariant theory 
\cite{GITconstructive}. This is essentially an algorithmic problem, albeit 
it can often be prohibitively intensive from a computational point of view. 

Another topic we have not discussed is the more general subject of 
{\em quiver field theories}. By this we mean supersymmetric field theories 
with a matter content chosen to transform in an arbitrary representation of a 
quiver (in the sense of \cite{quivers_intro, quivers_moduli}). 
Particular classes of such theories have been considered in \cite{AdSorb}, 
and they exhibit rather unusual properties. The mathematical theory of 
representations of quivers has been considered in \cite{quivers_moduli} 
and in the more recent mathematical literature \cite{quivers_recent}, 
and has deep connections 
with the representation theory of algebras. It would be interesting to 
understand the relevance of this relation from a physical point of view. 
Another subject worthy of consideration is the connection between such 
representations and noncommutative geometry, along the lines of 
\cite{nc}, as part of the more general philosophy according to which turning 
on Fayet-Iliopoulos terms is equivalent to making a noncommutative deformation
of the base space. The relevance of noncommutative geometry to matrix 
theory has been pointed out in a series of recent papers 
\cite{noncomm_geom}. 

On a more abstract level, it would be interesting to have a better 
understanding of the `location' of our brane-theoretic resolutions of 
singularities among the crepant resolutions guaranteed (for Gorenstein 
singularities of 3-folds) by the results of \cite{cansings}. That is, 
we would like to know which resolutions are realized by the D-brane mechanism 
out of the multitude of crepant resolutions which usually can 
be performed on a Gorenstein quotient singularity. This would give us a 
clearer picture of the way in which the D-brane effective field 
theory `projects out' certain 
phases which are otherwise permitted by classical geometry. 

Finally, one issue of major physical relevance is to what extent one can 
use our results and methods in order to extract information on the Maldacena 
limit of D-brane effective theories placed at various conical singularities 
which can be obtained from a quotient singularity by performing partial 
resolutions. A satisfactory answer to this question may provide a way of 
generalizing the work of \cite{MP} to the case of nonabelian quotient 
singularities, thus adding new strands to the web of connections between  
supersymmetric field theories, supergravities and string theory 
which is becoming increasingly apparent.

\bigbreak\bigskip\bigskip\centerline{{\bf Aknowledgements}}\nobreak
\bigskip

B.~R.~G. is supported by a National Young Investigator Award.  
The work of B.~R.~G. and C.~I~.L. is also supported by the DOE grant 
DE-FG02-92ER40699B.

\appendix

\section{The double symplectic quotient}

In order to justify our two-step procedure for taking the K\"{a}hler quotient, 
it suffices to consider the symplectic part of the problem (the fact that the 
complex structures agree is immediate). 

Suppose then that 
one has a symplectic manifold $(X,\omega)$ and a hamiltonian Lie group 
action on $(M,\omega)$ with (equivariant) moment map $M:X\rightarrow {\bf g}$, 
where ${\bf g}$ is the Lie algebra of $G$\footnote{We always use the Killing 
form $\langle,\rangle$ of ${\bf g}$ to identify it with ${\bf g}^*$.}. 
Picking a central Lie subgroup $T$ of $G$, its Lie algebra 
${\bf t}$ is a subalgebra of ${\bf g}$. By using the Killing form 
of ${\bf g}$, the Lie algebra ${\bf h}={\bf g}/{\bf t}$ of the quotient 
group $H:=G/T$ can be identified with the orthogonal complement of ${\bf t}$.
Let $P$ and $Q$ be the projectors on the orthogonal subspaces ${\bf t}$ and 
${\bf h}$ of ${\bf g}$ (these are orthoprojectors with respect to the Killing 
form of ${\bf g}$). 
The action of $T$ on $X$ induced from the action of $G$ is hamiltonian 
with moment map $M_T$ given by the projection of $M$ on ${\bf t}$:
\be
M_T=P\circ M:X\rightarrow {\bf t}~~.
\ee
Chosing a level $\xi \in {\bf t}$, one can consider the symplectic 
reduction ${\overline X}_T(\xi):=X_T(\xi)/T$ (with $X_T(\xi):=M_T^{-1}(\xi)$)
at level $\xi$, endowed with the induced symplectic form 
$\omega_T$. If $\pi_T:X_T(\xi)\rightarrow {\overline X}_T(\xi)$ is the natural 
projection, then $\omega_T$ is uniquely determined by the condition 
$\pi_T^*(\omega_T)=\omega|_{X_T(\xi)}$. Since $T$ is a central subgroup of 
$G$, we have an induced action of $H$ on ${\overline X}_T(\xi)$, which is 
easily seen to be hamiltonian with respect to $\omega_T$. Standard arguments 
show that its moment map is given by:
\be
M_H\circ \pi_T=Q\circ M:~M_T(\xi)\rightarrow {\bf h}~~.
\ee
Further reduction with respect to $H$ {\em at level zero} gives a manifold 
$\overline{{\overline X}_T(\xi)}_H(0):={\overline X}_T(\xi)_H(0)/H$, 
where ${\overline X}_T(\xi)_H(0):=M_H^{-1}(0)$. 
Denote the natural projection by 
$\pi_H:{\overline X}_T(\xi)_H(0)\rightarrow 
\overline{{\overline X}_T(\xi)}_H(0)$ and the induced symplectic form by 
$(\omega_T)_H$. 

On the other hand, one can consider the symplectic reduction 
${\overline M}_G(\xi)=M_G(\xi)/G$ ($M_G(\xi):=M^{-1}(\xi)$) of $M$ by $G$, 
with projection $\pi_G:=M_G(\xi)\rightarrow {\overline M}_G(\xi)$ and 
induced symplectic form $\omega_G$. One clearly has 
$\pi_T(M_G(\xi))={\overline M}_T(\xi)_H(0)$, which upon projecting by $\pi_H$ 
gives a diffeomorphism 
$\phi:{\overline M}_G(\xi)\rightarrow \overline{{\overline X}_T(\xi)}_H(0)$, 
which is easily seen to map $\omega_G$ into $(\omega_T)_H$. Hence we have a 
symplectic isomorphism:
\be
({\overline M}_G(\xi),\omega_G)\approx 
(\overline{{\overline X}_T(\xi)}_H(0),(\omega_T)_H)~~
\ee
for any level $\xi \in {\bf t}$.

\section{The method of projectors}

Consider a hermitian vector space $(W,<,>)$ carrying a unitary representation 
$\rho:\Gamma\rightarrow U(W,<,>)$ of a finite group $\Gamma$. 
Let $\rho_\mu$ ($\mu=0..r$) be the irreducible representations of $\Gamma$, 
and let $n_\mu$ be their dimensions. Pick a set of preffered 
realizations of these abstract irreducible 
representations by unitary matrices $D^{(\mu)}(\gamma) \in U(n_\mu)$.   

Consider the decomposition of $\rho$ into irreducible representations 
of $\Gamma$, realized by the orthogonal direct sum decomposition:
\be
W=\oplus_{\mu}{W_\mu}~~
\ee 
of $W$ into $\Gamma$-invariant subspaces, where the restriction of 
$\rho$ to each $W_\mu$ is equivalent to $a_\mu$ copies of $\rho_\mu$. 

Assuming that one knows the multiplicities $a_\mu$, and given an orthonormal 
basis $(e_s)_{s=1..n}$ of $W$, the method of projectors
allows one to determine fiducial orthonormal bases 
$(e^{(\mu,\alpha)}_i)_{i=1..n_\mu,\alpha=1..a_\mu}$ of $W_\mu$, i.e. 
orthonormal bases satisfying the condition:
$\rho(\gamma)e^{(\mu,\alpha)}_i=D^{(\mu)}_{ji}(\gamma)
e^{(\mu,\alpha)}_j$. 
In practice, the method produces the explicit expressions:
\be
e^{(\mu,\alpha)}_i=\sum_{j=1..n}{C^{(\mu i,\alpha)}_se_s}
\ee
of such basis elements in terms of the given basis $(e_s)_{s=1..n}$ of $W$. 
In particular, if $(W,\rho)$ is a tensor product of two irreducible 
representations $(R_\nu,\rho_\nu)$ and $(R_\lambda,\rho_\lambda)$ 
of $\Gamma$, and if one choses the basis of $W$ to be given by 
$e_{jk}:=e^{(\nu)}_j\otimes e^{(\lambda)}_k$, with $e^{(\nu)}_j$, 
$e^{(\lambda)}_k$ fiducial bases for $R_\nu,R_\lambda$, then the above 
expression becomes:
\be
e^{(\mu,\alpha)}_i=\sum_{j=1..n_\nu,k=1..n_\lambda}
{C^{(\mu i,\alpha)}_{\nu j,\lambda k}e^{(\nu)}_j\otimes e^{(\lambda)}_k}~~,
\ee
so that the method can be used to compute the Clebsch-Gordan coefficients 
$C^{(\mu i,\alpha)}_{\nu j,\lambda k}$.

For the general case of an arbitrary representation $(W,\rho)$, a set of 
fiducial basis vectors $e^{(\mu,\alpha)}_i$ can be determined as follows:

{\bf Step~1}: 
Define the linear operators:
\be
\label{ps}
P^{(\mu)}_{ji}:=\frac{n_\mu}{|\Gamma|}\sum_{\gamma\in \Gamma}
{D^{(\mu)}_{ji}(\gamma)^*\rho(\gamma)}~~(i,j=1..n_\mu)~~,
\ee
where $*$ denotes complex conjugation and $|\Gamma|$ is the order of $\Gamma$. 
The orthogonality relations for the matrices $D^{(\mu)}(\gamma)$ show that 
$P^{(\mu)}_{ji}$ satisfy the operator relations:
\bea
P^{(\mu)}_{kl} P^{(\nu)}_{ji} &=& \delta_{\mu\nu}\delta_{lj}P^{(\mu)}_{ki}~~\nn\\
\sum_{i}{P^{(\mu)}_{ii}}& = & id_{W}~~,
\eea
while unitarity of $D^{(\mu)}(\gamma)$ and $\rho(\gamma)$ imply:
\be
(P^{(\mu)}_{ji})^+=P^{(\mu)}_{ij}~~.
\ee
In particular, the operators $P^{(\mu)}_{ii}$ form a complete set of 
orthogonal projectors of $(W,<,>)$.

{\bf Step~2}: 
For each $\mu$ appearing in the decomposition of $W$ into irreducible 
representations, consider the space  
$W_1^{(\mu)}:=P^{(\mu)}_{11}(W)=ker P^{(\mu)}_{11}$ onto which 
$P^{(\mu)}_{11}$ projects. The dimensionality of this space coincides with 
the multiplicity 
$a_\mu$ of the irreducible representation $R_\mu$ in $W$. Compute an 
arbitrary orthonormal basis $(e^{(\mu\alpha)}_1)_{\alpha=1..a_\mu}$ 
of $W_1^{(\mu)}$.

{\bf Step~3}: 
Compute $e_i^{(\mu,\alpha)}:=P^{(\mu,\alpha)}_{i1}e^{(\mu,\alpha)}_1$, 
for each $\alpha=1..a_\mu$ and $i=2..n_\mu$. 
Then the vectors $(e^{(\mu,\alpha)}_i)_{\alpha=1..a_\mu,i=1..n_\mu}$  form 
a fiducial basis of the subspace $W_\mu$ (in particular, 
each subspace $W_\mu$ is determined as the span of these vectors).
 
The proof of the above statements consists of direct verifications and 
can be found for example in \cite{C_G}. In practice, the choice of a basis 
$(e_s)_{s=1..n}$ presents $W$ as the vector space $\C^n$ and identifies 
$e_s$ with the canonical basis vectors of $\C^n$. In this case, the 
representation $\rho$ is given by a set of $n$ by $n$ unitary matrices 
$\rho(\gamma)$ and the operators $P^{(\mu)}_{ji}$ given by (\ref{ps}) are 
identified with $n$ by $n$ matrices. Then the fiducial basis elements 
$e^{(\mu,\alpha)}_i$ produced by the above algorithm are realized as 
column vectors in $\C^n$. 
Since their expression in the canonical basis is 
simply given by their entries, it follows that $C^{(\mu i,\alpha)}_s$ 
is given by the entry number $s$ of the column vector $e^{(\mu,\alpha)}_i$.

\pagebreak

\end{document}

%% file: moment.pstex_t
\begin{picture}(0,0)%
\special{psfile=moment.pstex}%
\end{picture}%
\setlength{\unitlength}{0.006250in}%
\begingroup\makeatletter\ifx\SetFigFont\undefined
\def\x#1#2#3#4#5#6#7\relax{\def\x{#1#2#3#4#5#6}}%
\expandafter\x\fmtname xxxxxx\relax \def\y{splain}%
\ifx\x\y   
\gdef\SetFigFont#1#2#3{%
  \ifnum #1<17\tiny\else \ifnum #1<20\small\else
  \ifnum #1<24\normalsize\else \ifnum #1<29\large\else
  \ifnum #1<34\Large\else \ifnum #1<41\LARGE\else
     \huge\fi\fi\fi\fi\fi\fi
  \csname #3\endcsname}%
\else
\gdef\SetFigFont#1#2#3{\begingroup
  \count@#1\relax \ifnum 25<\count@\count@25\fi
  \def\x{\endgroup\@setsize\SetFigFont{#2pt}}%
  \expandafter\x
    \csname \romannumeral\the\count@ pt\expandafter\endcsname
    \csname @\romannumeral\the\count@ pt\endcsname
  \csname #3\endcsname}%
\fi
\fi\endgroup
\begin{picture}(535,310)(85,520)
\put(155,740){\makebox(0,0)[lb]{\smash{\SetFigFont{6}{7.2}{rm}$\phi$}}}
\put(375,645){\makebox(0,0)[lb]{\smash{\SetFigFont{6}{7.2}{rm}$\phi$}}}
\put(620,720){\makebox(0,0)[lb]{\smash{\SetFigFont{6}{7.2}{rm}$\phi$}}}
\put(535,520){\makebox(0,0)[lb]{\smash{\SetFigFont{6}{7.2}{rm}$\phi \phi^+ -\phi^+\phi$}}}
\put(125,520){\makebox(0,0)[lb]{\smash{\SetFigFont{6}{7.2}{rm}$-\phi^+\phi$}}}
\put(310,520){\makebox(0,0)[lb]{\smash{\SetFigFont{6}{7.2}{rm}$+\phi \phi^+$}}}
\put( 85,670){\makebox(0,0)[lb]{\smash{\SetFigFont{6}{7.2}{rm}$\mu$}}}
\put(155,815){\makebox(0,0)[lb]{\smash{\SetFigFont{6}{7.2}{rm}$\nu$}}}
\put(290,670){\makebox(0,0)[lb]{\smash{\SetFigFont{6}{7.2}{rm}$\mu$}}}
\put(405,560){\makebox(0,0)[lb]{\smash{\SetFigFont{6}{7.2}{rm}$\nu$}}}
\put(530,665){\makebox(0,0)[lb]{\smash{\SetFigFont{6}{7.2}{rm}$\mu$}}}
\end{picture}

%% file: orbits.pstex_t
\begin{picture}(0,0)%
\includegraphics{orbits.pstex}%
\end{picture}%
\setlength{\unitlength}{0.006250in}%
\begingroup\makeatletter\ifx\SetFigFont\undefined
\def\x#1#2#3#4#5#6#7\relax{\def\x{#1#2#3#4#5#6}}%
\expandafter\x\fmtname xxxxxx\relax \def\y{splain}%
\ifx\x\y   
\gdef\SetFigFont#1#2#3{%
  \ifnum #1<17\tiny\else \ifnum #1<20\small\else
  \ifnum #1<24\normalsize\else \ifnum #1<29\large\else
  \ifnum #1<34\Large\else \ifnum #1<41\LARGE\else
     \huge\fi\fi\fi\fi\fi\fi
  \csname #3\endcsname}%
\else
\gdef\SetFigFont#1#2#3{\begingroup
  \count@#1\relax \ifnum 25<\count@\count@25\fi
  \def\x{\endgroup\@setsize\SetFigFont{#2pt}}%
  \expandafter\x
    \csname \romannumeral\the\count@ pt\expandafter\endcsname
    \csname @\romannumeral\the\count@ pt\endcsname
  \csname #3\endcsname}%
\fi
\fi\endgroup
\begin{picture}(755,270)(40,530)
\put(580,745){\makebox(0,0)[lb]{\smash{\SetFigFont{6}{7.2}{rm}$H$}}}
\put(610,735){\makebox(0,0)[lb]{\smash{\SetFigFont{6}{7.2}{rm}$T$}}}
\put(305,740){\makebox(0,0)[lb]{\smash{\SetFigFont{6}{7.2}{rm}$H$}}}
\put(575,655){\makebox(0,0)[lb]{\smash{\SetFigFont{6}{7.2}{rm}
$T^\C$
}}}
\put(705,585){\makebox(0,0)[lb]{\smash{\SetFigFont{6}{7.2}{rm}
${\cal O}$
}}}
\put(705,725){\makebox(0,0)[lb]{\smash{\SetFigFont{6}{7.2}{rm}
${\cal O'}$
}}}
\put(640,530){\makebox(0,0)[lb]{\smash{\SetFigFont{6}{7.2}{rm}
${\cal Z}_\xi$
}}}
\put(280,780){\makebox(0,0)[lb]{\smash{\SetFigFont{6}{7.2}{rm}
$P_{\xi'}$
}}}
\put( 40,780){\makebox(0,0)[lb]{\smash{\SetFigFont{6}{7.2}{rm}
${\cal M}_{\xi'}$
}}}
\put(600,785){\makebox(0,0)[lb]{\smash{\SetFigFont{6}{7.2}{rm}
${\cal Z}_{\xi'}$
}}}
\put(290,530){\makebox(0,0)[lb]{\smash{\SetFigFont{6}{7.2}{rm}
$P_\xi$
}}}
\put( 40,530){\makebox(0,0)[lb]{\smash{\SetFigFont{6}{7.2}{rm}
${\cal M}_\xi$
}}}
\put(265,670){\makebox(0,0)[lb]{\smash{\SetFigFont{6}{7.2}{rm}
$\Psi$
}}}
\end{picture}